\documentclass[aps,twocolumn,showpacs,superscriptaddress,floatfix]{revtex4-1}
\usepackage{amsfonts}

\usepackage{graphicx}
\usepackage{amsmath}
\usepackage{amssymb}
\usepackage{bm}
\usepackage{gensymb}

\usepackage{hyperref}
\hypersetup{
        colorlinks=true,
        urlcolor=blue,
      	colorlinks=false,
      	pdfborder={0 0 0},
}

\usepackage{float}

\newcommand{\lb}{\langle}
\newcommand{\rb}{\rangle}

\usepackage{xcolor}
\definecolor{fqColor}{rgb}{1, 0.65, 0.48}
\definecolor{afqColor}{rgb}{1, 0.45, 0.5} 
\definecolor{fmColor}{rgb}{0, 0.12, 1} 
\definecolor{afm3Color}{rgb}{0.5, 0, 0.5}

\begin{document}
\title{A ground state study of the spin-1 bilinear-biquadratic Heisenberg model on the triangular lattice using tensor networks}

\author{Ido Niesen}
\affiliation{Institute for Theoretical Physics and Delta Institute for Theoretical Physics, University of Amsterdam, Science Park 904, 1098 XH Amsterdam, The Netherlands}

\author{Philippe Corboz}
\affiliation{Institute for Theoretical Physics and Delta Institute for Theoretical Physics, University of Amsterdam, Science Park 904, 1098 XH Amsterdam, The Netherlands}

\date{\today}

\begin{abstract}
Making use of infinite projected entangled pair states, we investigate the ground state phase diagram of the nearest-neighbor spin-1 bilinear-biquadratic Heisenberg model on the triangular lattice. In agreement with previous studies, we find the ferromagnetic, 120$\degree$ magnetically ordered, ferroquadrupolar and antiferroquadrupolar phases, and confirm that all corresponding phase transitions are first order. Moreover, we provide an accurate estimate of the location of the ferroquadrupolar to 120$\degree$ magnetically ordered phase transition, thereby fully establishing the phase diagram. Also, we do not encounter any signs of the existence of a quantum paramagnetic phase. In particular, contrary to the equivalent square lattice model, we demonstrate that on the triangular lattice the one-dimensional Haldane phase does \emph{not} reach all the way up to the two-dimensional limit. Finally, we investigate the possibility of an intermediate partially-magnetic partially-quadrupolar phase close to $\theta = \pi/2$, and we show that, also contrary to the square lattice case, this phase is not present on the triangular lattice. 
\end{abstract}



\maketitle

\section{Introduction}

	Geometric frustration in strongly correlated materials can cause even relatively simple systems to develop unexpected types of order. For magnetic materials, the archetypal example of a geometrically frustrated system is the triangular lattice Heisenberg antiferromagnet. In 1973, Anderson~\cite{anderson73} proposed that the spin-1/2 triangular antiferromagnet has a  ground state consisting of resonating valence bonds, also called a quantum spin liquid. However, it was later shown numerically~\cite{bernu92,capriotti99} that the ground state is ordered instead and displays $120$-degree magnetic order. 

	In this paper, we focus on the two-dimensional triangular lattice spin-1 Heisenberg model with additional biquadratic coupling|known as the bilinear-biquadratic Heisenberg (BBH) model. It is defined by the following Hamiltonian
	\begin{equation}
		H = \sum_{\lb i,j \rb} \cos(\theta) \, \bm{S}_i \cdot \bm{S}_j + \sin(\theta) \left(\bm{S}_i \cdot \bm{S}_j\right)^2,
		\label{Ham}
	\end{equation}
	where the sum goes over all nearest neighbors, $\bm{S}_i = (S^x_i, S^y_i, S^z_i)$ is the vector of spin-matrices for the spin-1 particle on site $i$, and $\theta \in [0,2\pi)$ determines the strength of the biquadratic term relative to the bilinear term. The biquadratic term can appear as a second order correction in the expansion of the exchange interaction, in which case it is small compared to the bilinear term. However, it has been argued that a significant biquadratic interaction may exist. For example,  the behavior of the magnetic susceptibility of the one-dimensional material LiVGe$_2$O$_6$ can be explained by a significant biquadratic interaction~\cite{millet99}, a suggested underlying microscopic mechanism of which can be found in Ref.~\cite{mila00}.

	The triangular lattice spin-1 BBH model gained attention recently because it was suggested that both its antiferroquadrupolar~\cite{tsunetsugu06,tsunetsugu07,li07} and ferroquadrupolar~\cite{bhattacharjee06,stoudenmire09,nakatsuji10} ground state phases could give a possible explanation for the unusual behavior~\cite{nakatsuji05,nakatsuji07,nakatsuji10} of NiGa$_2$S$_4$. 	Moreover, Cheng et al.~\cite{cheng11} found spin-liquid-like behavior of the 6$H$-$B$ phase of the two-dimensional triangular magnet Ba$_3$NiSb$_2$O$_9$~\cite{fak17,xu12}, for which Serbyn et al.~\cite{serbyn11} proposed a candidate spin-liquid ground state that within the mean-field approximation was supposed to be a ground state the triangular spin-1 BBH model with additional single-ion anisotropy. However, a further investigation by Bieri et al.~\cite{bieri12} demonstrated that the the spin-liquid state found by Serbyn et al.~\cite{serbyn11} turned out \emph{not} to be the lowest energy state of the triangular spin-1 BBH model with single ion anisotropy.

	Additionally, at $\theta = \pi/4$, the BBH model is equivalent to the SU(3) Heisenberg model, which could potentially be simulated using cold atoms trapped in an optical lattice~\cite{imambekov03,wu03,honerkamp04,cazalilla09,becker10,gorshkov10,struck13,jo12}. Besides, as the most general lattice-translation, lattice-rotation and spin-rotation-symmetric spin-1  Hamiltonian with nearest-neighbor interactions, the BBH Hamiltonian is interesting in its own right from a theoretical point of view.

	Moreover, in our recent study of the spin-1 BBH model on the square lattice~\cite{niesen17,niesen17b}, we found the occurrence of a quantum paramagnetic phase in between the antiferromagnetic and 120$\degree$ magnetically ordered phases, and we were able to show that this phase can be adiabatically connected to the one-dimensional Haldane phase of decoupled spin-1 chains. In addition, we also encountered a partially-magnetic partially-quadrupolar phase in between the antiferroquadrupolar and ferromagnetic phases. Both discoveries raise the question whether any of the above phenomena also manifest themselves on the experimentally more relevant triangular lattice.

	Finally, it should be noted that the spin-1 BBH model on the triangular lattice is very challenging to study from a numerical perspective, as it suffers from the negative sign problem. Previous studies|summarized in Section~\ref{sec-prevst}|are either based on approximate methods, or involve exact diagonalization on small systems; however, a complete and systematic study is still lacking.

\begin{figure}[htb]
	\begin{center}
	\includegraphics[scale=0.42]{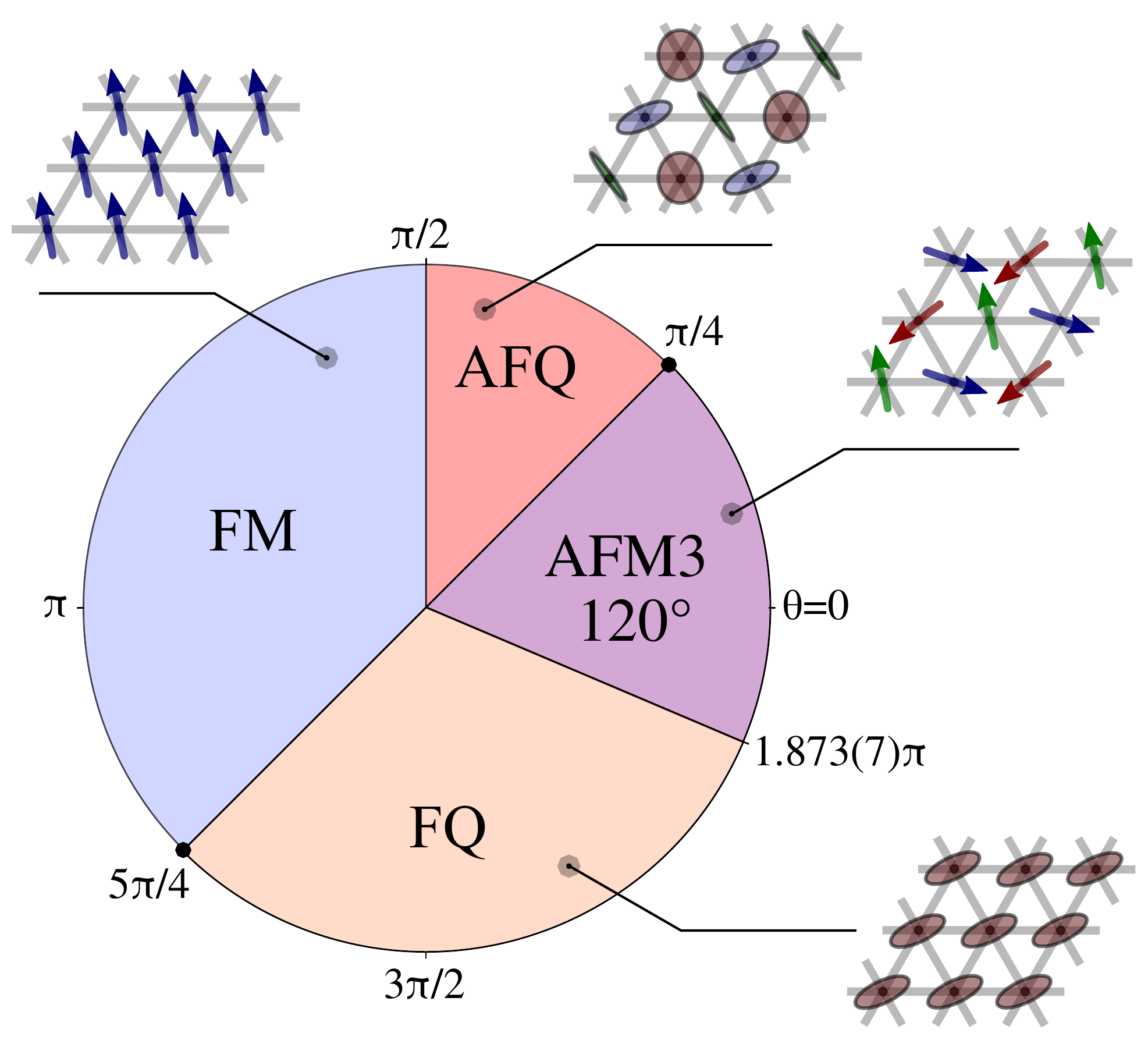}
	\end{center}
	\caption{The iPEPS ground state phase diagram. In anti-clockwise order starting at $\theta = 0$, we have the 120 degree magnetically ordered (AFM3), antiferroquadrupolar (AFQ), ferromagnetic (FM) and ferroquadrupolar (FQ) phases. The SU(3)-symmetric points at $\theta = \pi/4$ and $5\pi/4$ are labeled by black dots.}
	\label{fig-ipeps-pd}
	\end{figure}

	In this work, we use state-of-the-art tensor network algorithms to map the entire ground state phase diagram of the triangular lattice spin-1 BBH model|displayed in Fig.~\ref{fig-ipeps-pd}. In agreement with previous studies, we find the ferromagnetic (FM), ferroquadrupolar (FQ), 120$\degree$ antiferromagnetically ordered (AFM3)~\cite{n-threesubl} and antiferroquadrupolar (AFQ) phases and obtain an accurate estimate of the location of the FQ to AFM3 transition|which we predict to occur at $\theta_c = 1.873\pi  \pm 0.007 \pi$|thereby fully establishing the phase diagram.

	Motivated by our findings for the corresponding square lattice model~\cite{niesen17,niesen17b}, we also investigate the anisotropic triangular lattice spin-1 BBH model for $\theta \in (-\pi/4,\pi/4)$. We show that, on the triangular lattice, the one-dimensional Haldane phase does \emph{not} reach up to the two-dimensional isotropic model|albeit that it \emph{does} extend far from the one-dimensional limit. In addition, the triangular-lattice equivalent of the partially-magnetic partially-quadrupolar phase found on the square lattice is shown to \emph{not} be present in the ground state phase diagram of the triangular lattice spin-1 BBH model.

	This paper is organized as follows. After going over some general background information on the spin-1 BBH model in Section~\ref{sec-prelim}, we set the stage by discussing previous work that has been done on the triangular lattice spin-1 BBH model in Section~\ref{sec-prevst}, and identify possible points of interest. We then provide an overview of the numerical method used in Section~\ref{sec-method}, and present our results concerning the ground state phase diagrams of the isotropic and anisotropic triangular lattice spin-1 BBH models in Section~\ref{sec-results}. Finally, we discuss our findings in Section~\ref{sec-conclusion}.

\section{Model}
\label{sec-model}

\subsection{Preliminaries}
\label{sec-prelim}

	The biquadratic part $(\bm{S}_i \cdot \bm{S}_j)^2$ of the Hamiltonian~(\ref{Ham}) introduces on-site products of spin operators $S^\alpha S^\beta$, where $\alpha,\beta \in \{x,y,z\}$. Thought of as a 3x3 matrix (with indices $\alpha$ and $\beta$), the trace and the anti-symmetric parts of $S^\alpha S^\beta$ are just (a multiple of) the total spin $S=1$ (times the identity matrix) and the original spin dipole vector $\bm{S}$ (due to the spin-commutation relations) respectively. What remains is the traceless symmetric part of $S^\alpha S^\beta$, which is called the \emph{quadrupole matrix}, or Q-matrix for short,
	\begin{equation*}
	 	Q^{\alpha \beta} := S^\alpha S^\beta + S^\beta S^\alpha - \frac{2}{3}S(S+1)\delta^{\alpha \beta}.
	\end{equation*}
	The five independent components of $Q^{\alpha \beta}$ can be organized into a single vector, denoted by boldface $\bm{Q}$, as follows
	\begin{equation*}
	    \bm{Q} := 
	    \left(\begin{matrix}
	    (S^x)^2 - (S^y)^2 \\ 
	    \frac{1}{\sqrt{3}}\left[2(S^z)^2 - S(S+1)) \right] \\ 
	    S^x S^y + S^y S^x \\
	    S^y S^z + S^z S^y \\
	    S^z S^x + S^x S^z 
	    \end{matrix} \right).
	\end{equation*}
	Rewriting the Hamiltonian in (\ref{Ham}) in terms of the quadrupolar vectors $\bm{Q}_i$ (see~\cite{penc11}) yields 	
	\begin{equation}
	    H = \sum_{\lb i,j \rb} J_S(\theta) \, \bm{S}_i \cdot \bm{S}_j + J_Q(\theta) \, \bm{Q}_i \cdot \bm{Q}_j,
		\label{Ham_quad}
	\end{equation}
	up to an additive $\theta$-dependent constant of $4\sin(\theta) /3$ that is not relevant for this discussion. The spin and quadrupolar coupling constants are given by $J_S(\theta) = \cos(\theta) - \sin(\theta)/2$ and $J_Q(\theta) = \sin(\theta)/2$.

	The advantage of expressing the Hamiltonian in terms of $\bm{S}$ and $\bm{Q}$ (\ref{Ham_quad}) over the original notation (\ref{Ham}), is that the former clearly separates the dipolar and quadrupolar terms|related to magnetic and (spin) \emph{nematic} order respectively. Nematic order involves the breaking of spin-rotational symmetry while preserving time-reversal symmetry~\cite{andreev84}; it differs from magnetic order in that the latter breaks both spin-rotation and time-reversal symmetry. We will provide examples of both types of order in Section~\ref{sec-prevst}. Moreover, as described in for example Ref.~\cite{niesen17b}, the points of enhanced SU(3)-symmetry are made explicit in~(\ref{Ham_quad}), which in the case of the tripartite triangular lattice are  those for which $J_S = J_Q$, i.e. $\theta = \pi/4$ and $5\pi/4$.
	
	Technically, quadrupolar order is described by the spectrum of the Q-matrix. Since $\text{tr}(Q) = 0$, the spectrum of Q is fully determined by two matrix invariants, for which there are many possible choices, such as: two out of three eigenvalues, or, the invariants $II_Q = -\frac{1}{2}\text{tr}(Q)^2$ and $III_Q = \text{det}(Q)$ used in~\cite{niesen17b}. However, when we are searching for jumps in the Q-matrix spectrum that signify a (first order) phase transition, finding a jump in just one matrix invariant is sufficient. An obvious choice for this one invariant is the vector norm $|\bm{Q}| = \sqrt{\bm{Q} \cdot \bm{Q}}$, which~\cite{penc11} is also equal to ($1/\sqrt{2}$ times) the Frobenius norm $\sqrt{\text{tr}(Q^{\dagger} Q)}$ of the Q-matrix. We will refer to this norm as the \emph{Q-norm} for short.

\subsection{Previous studies}
\label{sec-prevst}

	Spin-1 BBH models have been extensively studied throughout the years. Of relevance to our investigation of the triangular lattice BBH model in particular, is the pioneering construction of an exact ground state~\cite{affleck87} of the spin-1 BBH chain at $\theta = \arctan(1/3)$ in the middle of the Haldane phase~\cite{haldane83,haldane83b}, and the development of a form of spin-wave theory by Papanicolaou~\cite{papanicolaou88} that is readily applicable to spin-1 BBH models on bipartite lattices. However, it was not until 2005 that a discovery of a low temperature spin-disordered state in the triangular magnet NiGa$_2$S$_4$ by Nakatsuji et al.~\cite{nakatsuji05,nakatsuji07}) sparked a surge of increased interest in the spin-1 BBH model on the two-dimensional triangular lattice, initializing a series of vibrant discussions on the nature of this new-found spin disordered state. 

\begin{figure}[htb]
	\begin{center}
	\includegraphics[scale=0.4]{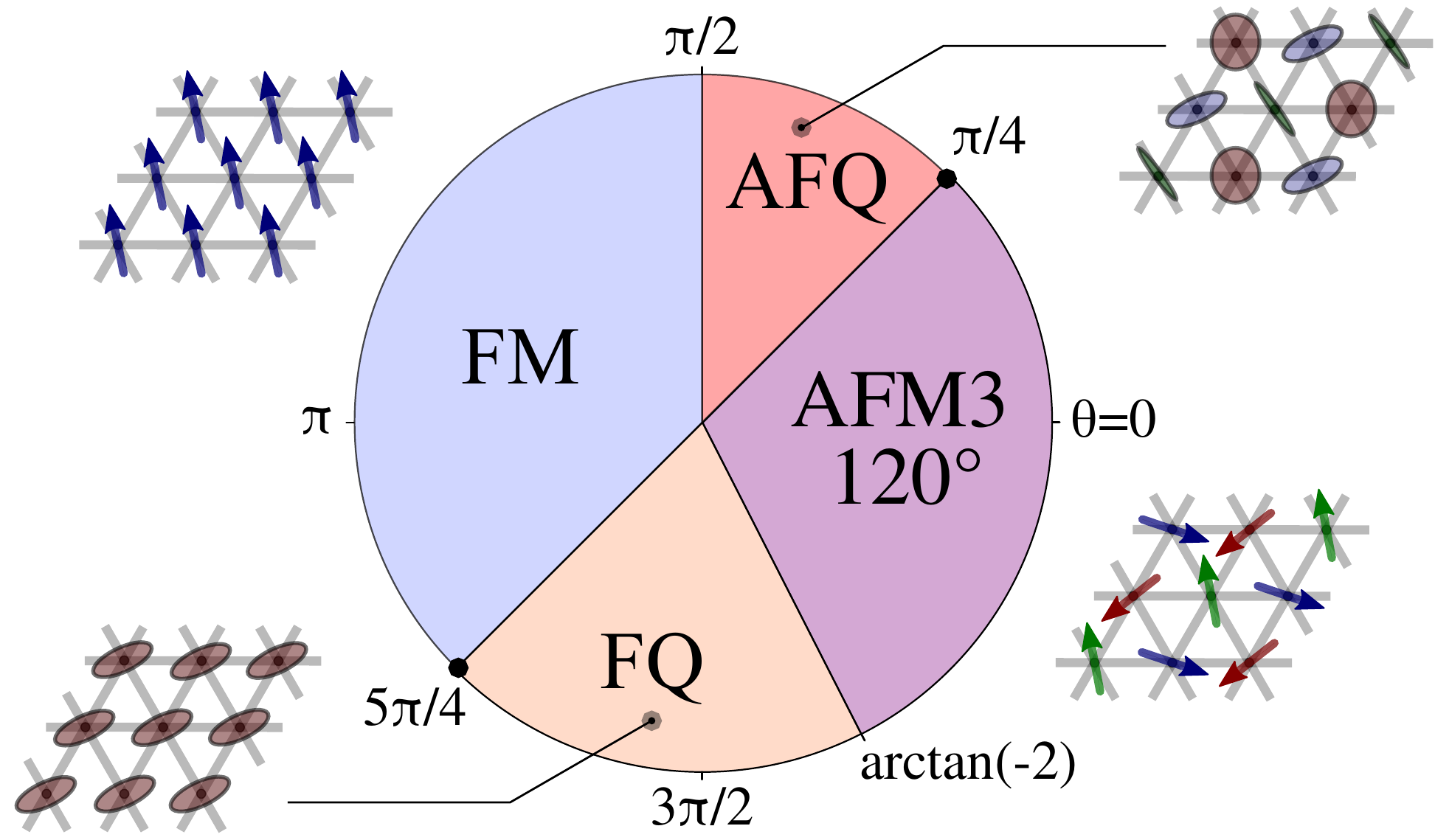}
	\end{center}
	\caption{The product ground state phase diagram. The FQ to AFM3 phase transition occurs at $\Theta_c^{MF} = \text{arctan}(-2) \approx 1.648 \pi$ (L\"{a}uchli et al.~\cite{lauchli06}).}
	\label{fig-product-gs}
	\end{figure}

	In order to introduce some terminology, let us first discuss the already very rich product ground state phase diagram of the triangular lattice spin-1 BBH model|shown in Fig.~\ref{fig-product-gs}|computed by L\"{a}uchli et al.~\cite{lauchli06} assuming a tripartite site-factorized product state ansatz for the ground state. 
	In addition to the magnetized 120$\degree$ antiferromagnetically ordered (AFM3) and ferromagnetic (FM) phases, Fig.~\ref{fig-product-gs} also contains the nematic ferroquadrupolar (FQ) and antiferroquadrupolar (AFQ) phases. 

	States in any of the quadrupolar phases have zero on-site magnetic dipole moment $\lb \bm{S} \rb = 0$ (since time-reversal symmetry is preserved), but they break spin-rotation symmetry by developing an anisotropy in their spin fluctuations. A typical example of a quadrupolar single-particle state is the $|0\rb$ state in the $S^z$-basis. This state clearly satisfies $\lb 0 |\bm{S} | 0 \rb = 0$, but it breaks spin-rotation symmetry because $\lb 0 | (S^z)^2 | 0 \rb = 0$ whereas $\lb 0 | (S^{x})^2 | 0 \rb = 1 = \lb 0 | (S^{y})^2 | 0 \rb$, indicating that the spin-vector fluctuates only in the x-y plane. A unit vector perpendicular to the plane of fluctuations ($\pm \bm{e}_z$ in the case of $|0\rb$) is called a \emph{director}. Now, product states in the ferroquadrupolar phase have directors on neighboring sites align in the same direction, whereas states in the antiferroquadrupolar phase have neighboring directors align in mutually perpendicular directions|e.g. in the $x$, $y$ and $z$-directions|assuming a three-sublattice pattern. The quadrupolar states are pictured by discs in Fig.'s~\ref{fig-ipeps-pd} and~\ref{fig-product-gs} representing their planes of spin fluctuations.

	In the same paper~\cite{lauchli06}, by means of exact diagonalization, L\"{a}uchli et al.~continued to show that the mean-field critical parameter value $\Theta_c^{MF} = \arctan(-2)$  separating the FQ and AFM3 phases gets renormalized to $\Theta_c^{ED} \approx -0.11\pi$. Moreover, L\"{a}uchli et al.~also found the occurrence of a $m=2/3$ magnetization plateau in the AFQ phase without the occurrence of a $m=1/3$ magnetization plateau (something that is unlikely to occur for purely magnetic states because of the lower commensurability of the latter), and they consider this a characteristic of the AFQ phase. 

	Following the above-mentioned discovery of Nakatsuji et al.~\cite{nakatsuji05},
	Tsunetsugu and Arikawa~\cite{tsunetsugu06,tsunetsugu07} proposed that the AFQ phase of the triangular lattice spin-1 BBH model could explain many features of the new-found spin disordered state in NiGa$_2$S$_4$. However, Bhattacharjee et al.~\cite{bhattacharjee06} and later also Nakatsuji et al.~\cite{nakatsuji10} both suggested that instead the FQ phase is a more likely candidate for the same new-found spin-disordered state. Additionally, there have been several Monte Carlo studies of the triangular lattice spin-1 BBH model concerning the pure biquadratic point $\theta = -\pi/2$~\cite{kaul12} and the $\theta \in [\pi, 3\pi/2]$ part~\cite{voll15}|both of which are in agreement with the phase diagram from L\"{a}uchli et al.~\cite{lauchli06}.

	More recently, we conducted an iPEPS study of the spin-1 BBH model on the two-dimensional square lattice~\cite{niesen17,niesen17b}, which yielded the occurrence of two additional phases on top of those present in the product ground state phase diagram. In particular, we found that in between the ordinary antiferromagnetic (AFM) and 120$\degree$ magnetically ordered (AFM3) phases, a quantum paramagnetic phase arises~\cite{niesen17} that is characterized by the fact that it preserves spin-rotation and lattice-translation symmetry while breaking lattice-rotation symmetry due to energy differences between the $x$ and $y$-bonds. Moreover, by continuously shrinking all high energy bond couplings to zero, the quantum paramagnetic phase turned out to be adiabatically connected to the Haldane phase of decoupled spin-1 chains, and can thus be viewed as a two-dimensional extension of the latter. 

	The AFM to Haldane transition, which was first predicted in Ref.~\cite{read89}, has also been the subject of related works on the square lattice spin-1 $J_1$-$J_2$ model~\cite{jiang09,haghshenas18b}, the square lattice next-nearest-neighbor spin-1 BBH $J_1$-$J_2$-$K_1$-$K_2$ model~\cite{gong17} and the $J_1$-$J_2$-$K_1$ model~\cite{chen18}. Finally, Lee and Kawashima~\cite{lee18}  also found a spin-liquid-like ground state on the spin-1 BBH model on the star lattice in a parameter regime that encloses the region for which we found the Haldane phase on the square lattice.

	In addition to the Haldane phase, on the square lattice we also encountered~\cite{niesen17b} the $m=1/2$ partially-magnetized  partially-quadrupolar phase|a phase that was predicted to appear only in the presence of an external magnetic field by T\'{o}th et al.~\cite{toth12}|and found that this phase is also present in the zero-field phase diagram. 

	Motivated by the above discoveries, we shall investigate the region $-\pi/4 < \theta < \pi/4$ where the Haldane phase occurs in the one-dimensional BBH chain, and keep an eye out for possible intermediate quantum paramagnetic phases. Moreover, in light of the characteristic $m=2/3$ magnetization plateau in the AFQ phase in the presence external magnetic field~\cite{lauchli06}, we will also investigate the possibility that the $m=2/3$ phase extends to the zero-field phase diagram of the triangular lattice spin-1 BBH model.

\section{Method}
\label{sec-method}

\subsection{iPEPS algorithm}

	We simulate the ground state of the two-dimensional triangular lattice spin-1 BBH model in the thermodynamic limit using a variational tensor network ansatz called an \emph{infinite projected entangled pair state}, or iPEPS \cite{jordan08,corboz10} for short. iPEPS is the infinite-system version of PEPS~\cite{verstraete04,nishino01,nishio04}: a two-dimensional generalization of matrix product states (MPS)~\cite{fannes92,ostlund95,rommer97}, the latter being the ansatz underlying the well-known density-matrix-renormalization-group (DMRG) algorithm~\cite{white92,white93,schollwock05}. 

\begin{figure}[htb]
	\includegraphics[scale=0.4]{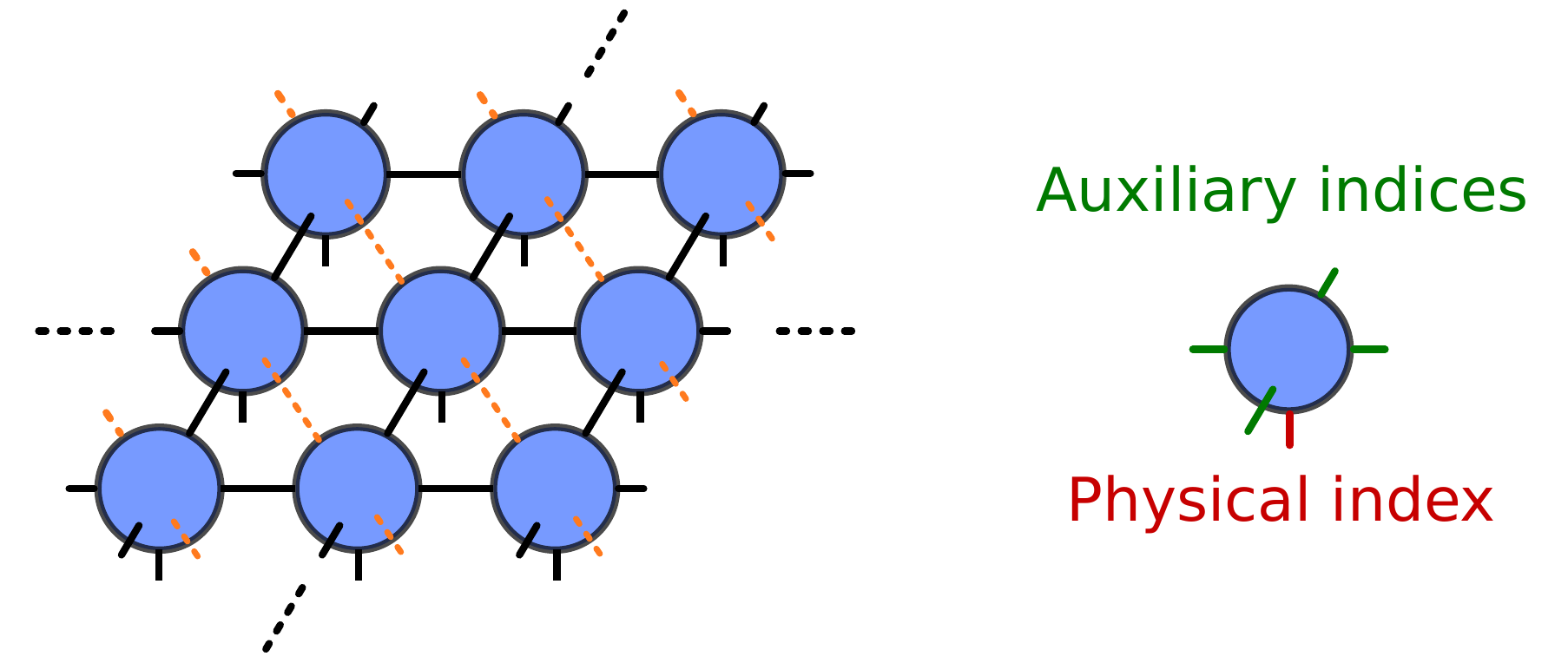}
	\caption{The iPEPS ansatz for the ground state. Each site is represented by a five-legged tensor (right), with four auxiliary indices and one physical index. The orange lines (left) are \emph{not} present in the actual iPEPS, but symbolically represent the diagonal interactions present only in the triangular lattice Hamiltonian.}
	\label{fig-ipeps}
	\end{figure}

	The iPEPS we use in our simulations consist of five-legged tensors, one per site, with one physical leg corresponding to the spin-1 particle on the site in question, and four auxiliary legs that connect to neighboring tensors forming a square lattice pattern. The triangular lattice structure of the model is not encoded in the iPEPS, but only in the Hamiltonian, which contains additional interaction terms along the diagonal lines (Fig.~\ref{fig-ipeps}).
	Each iPEPS is defined by a given unit cell of tensors|depending on if and how the ground state breaks translational symmetry|that is repeated all over the lattice.

	The vector spaces corresponding to the auxiliary legs all have the same dimension|called the \emph{bond dimension}, denoted by $D$|that controls the accuracy of the iPEPS. $D=1$ describes product states, and as $D$ increases, more entanglement can be encoded in the iPEPS. In the $D\rightarrow \infty$ limit the iPEPS can theoretically describe any state in the Hilbert space. Therefore, when computing expectation values, we will take the $D \rightarrow \infty$ limit when precise quantities are needed (and the ground state is not a product state).

	Computing expectation values requires contracting an infinite tensor network, which in two dimensions can only be done approximately. We use a variant~\cite{corboz14,corboz11b} of the corner-transfer matrix (CTM) algorithm~\cite{nishino96,nishino97,nishino00,orus09} based on a formalism derived by Baxter~\cite{baxter68,baxter78}. This requires the introduction of an additionally \emph{boundary bond dimension} $\chi$. In practice, we take $\chi(D) > D^2$ to be large enough that the error due to the use of finite $\chi$ is negligible compared to the error due to the use of finite $D$.

	Given an initial iPEPS, we obtain a low energy state by evolving it in imaginary time using a triangular lattice variant of either the simple~\cite{jiang08} or the more accurate (but computationally more expensive) full update algorithm~\cite{jordan08,corboz10,phien15}. The triangular lattice simple update algorithm is a modified version of the simple update method for square lattice Hamiltonians with an additional next-nearest neighbor interaction from Ref.~\cite{corboz10b}. The difference lies in the fact that, instead of truncating the bond dimension back to $D$ immediately after applying a \emph{single} imaginary time evolution gate, the triangular lattice algorithm simultaneously applies a horizontal, vertical, and diagonal evolution gate, and only afterwards truncates the bond dimensions back to $D$.

	The triangular lattice full update method used here is a variant of the next-nearest neighbor method from Ref.~\cite{corboz13} (see also Ref.~\cite{haghshenas18b}). After time-evolving a given iPEPS $|\Psi \rb$ a small step in imaginary time by means of the evolution gate $g$~\cite{n-itegate}|which increases the bond dimension|the optimal time-evolved iPEPS $|\tilde{\Psi}\rb$ with bond dimension $D$ is obtained by minimizing the norm distance $|| g\Psi - \tilde \Psi ||$. In the regular full update this is usually done by iteratively minimizing over two tensors $(p,q)$ on a bond until the cost function $|| g\Psi - \tilde \Psi ||$ has converged (cf. Refs.~\cite{corboz10,phien15}). In the presence of an additional diagonal interaction we need to optimize over four of these tensors, two on a horizontal bond ($p_h$,  $q_h$) and two on a vertical bond ($p_v$, $q_v$), respectively. We do this by performing an outer loop where we switch between the horizontal and vertical pairs of tensors, and an inner loop where we iteratively optimize over the corresponding pair of tensors.

	Where necessary, we do additional checks using the variational update algorithm~\cite{corboz16b} generalized to next-nearest neighbor interactions, which for fixed D gives the best results, but can as of yet not always be pushed to as high a bond dimension as the simple and full update simulations can be.  

	For a given value of $\theta$, the ground state of the system is the imaginary-time-evolved or variationally-optimized iPEPS with the lowest energy of the different unit cells considered. In this work, we have used unit cells consisting of up to $6 \times 6$ sites.

	We can choose to start a simulation either from a randomly initialized iPEPS, or from an iPEPS that is already in a certain phase. Making use of hysteresis, the latter can be particularly useful to determine the critical value of $\theta$ for which a given (first order) phase transition occurs, which we have done for the ferroquadrupolar to 120$\degree$ magnetically ordered phase transition.

\subsection{SU(3)-symmetric point benchmark}
\label{sec-benchmk}

\begin{figure}[htb]
	\includegraphics[scale=0.5]{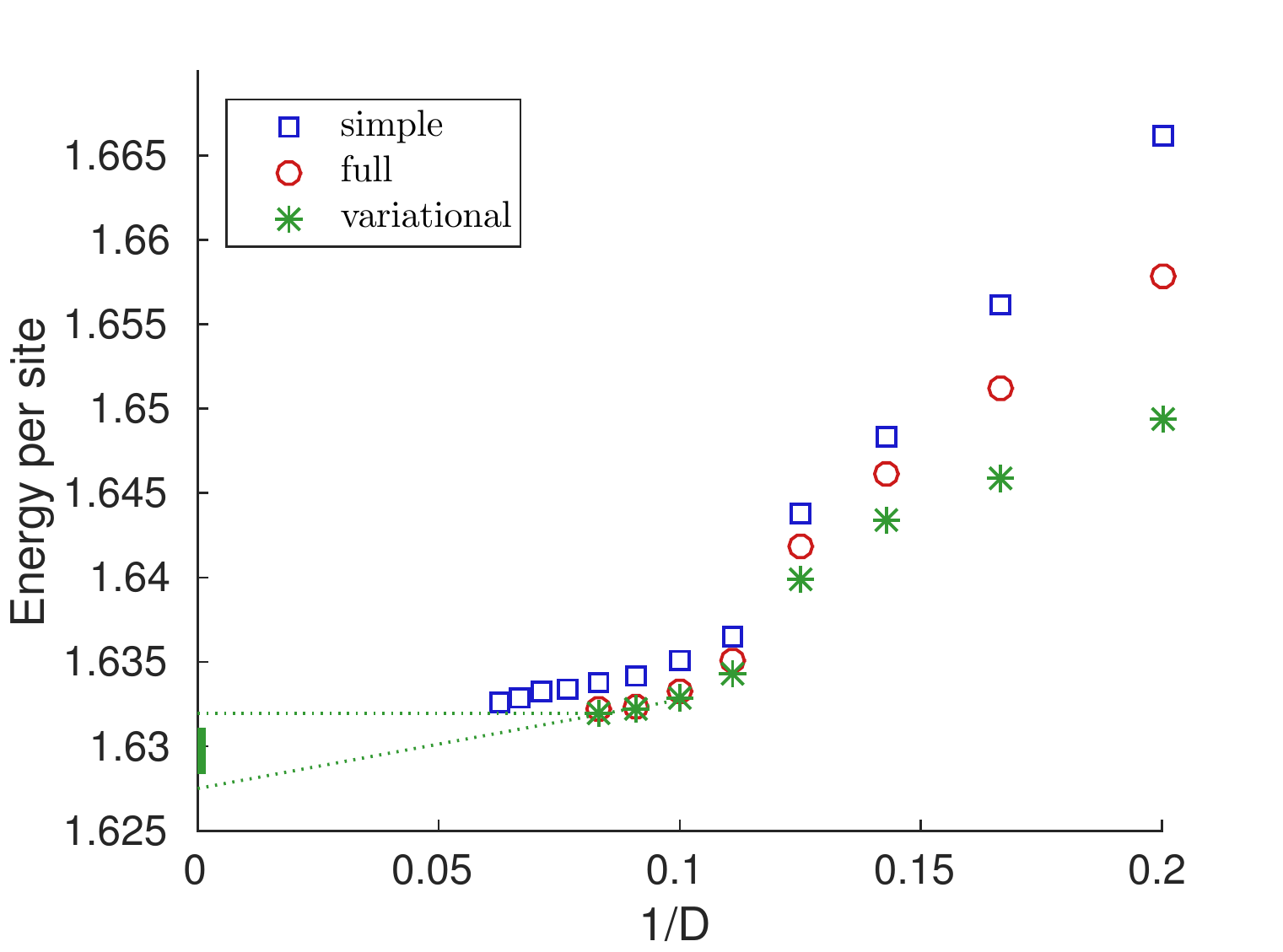}
	\caption{Energy per site at the SU(3) point $\theta = \pi/4$ for the simple, full and variational update. The simulations make use of the SU(3) symmetry of the ground state and can therefore be ran at higher bond dimension than usual. We take the midpoint of the ends of the two dotted lines to be the $D\rightarrow \infty$ extrapolated energy per site (see main text).}
	\label{fig-su3-bm}
	\end{figure}

	Before we proceed, let us benchmark the triangular lattice simple, full and variational update algorithms at the SU(3)-symmetric point $\theta = \pi/4$ by comparing to a previous study of the SU(3)-Heisenberg model by Bauer et al.~\cite{bauer12}.  We make use of the additional symmetries of the Hamiltonian to push the bond dimension to $D=16$ for the simple update, and $D=12$ for the full and variational updates. (See Refs.~\cite{singh11,bauer11} on how to implement global abelian symmetries within the tensor network formalism.) The resulting energies per site are shown in Fig.~\ref{fig-su3-bm}.

	Fig.~\ref{fig-su3-bm} shows that the full and variational update both give a visible improvement over the simple update. This reflects the fact that the former two both use the full environment at each optimization step, whereas the simple update only uses an approximate environment|making it computationally cheaper.
	
	Because the ansatz used for the ground state is variational, each iPEPS energy computed is higher than or equal to the true ground state energy. Therefore, the lowest finite-$D$ energy obtained|in this case given by the $D=12$ variational update simulation (shown in green) with an energy of $E^{\text{var}}_{D=12} = 1.632$|serves as an upper bound for the true ground state energy. Since increasing $D$ introduces more variational parameters in the iPEPS, the energy of the iPEPS decreases as $D$ increases. However, the behavior of the energy of an iPEPS is typically such that the energy as a function of $1/D$ curve flattens out as $D$ increases. Thus, a lower bound for the true ground state energy can be obtained by drawing a straight line through the last few high-$D$ data points (depicted by the lowest dotted green line in Fig.~\ref{fig-su3-bm}) and extrapolating it to $D \rightarrow \infty$. We shall take the average of the lowest obtained finite-$D$ energy ($E^{\text{var}}_{D=12}$) and the straight-line extrapolated energy through the last few data points as our estimate for the true ground state energy, which in this case yields $E^{\text{var}}_{D \rightarrow \infty} = 1.630$. Because the above-mentioned bounds are loose bounds, for our estimate of the error we shall take half of the difference between the lowest obtained energy and the straight-line extrapolated energy, resulting in an extrapolated energy of $E^{\text{var}}_{D \rightarrow \infty} = 1.630(1)$. The error bar is depicted by the thin green slab on the y-axis in Fig.~\ref{fig-su3-bm}~\cite{n-errorbar}.

 	Contrasted to the result obtained by Bauer et al.~\cite{bauer12} of $E_{D \rightarrow \infty}^{\text{previous}} = 1.633(14)$, we can conclude that our result is not only slightly lower in energy, but also more accurate; in part because we can go to higher bond dimension, but also  due to algorithmic improvement. Indeed, the lowest finite-$D$ energy obtained by Bauer et al.~is a $D=10$ simulation with an energy of $E^{\text{previous}}_{D=10} = 1.646$, which is higher than our $D=10$ simple update energy. Note that the ground state energy per site of the SU(3) Heisenberg model is related to the ground state energy per site of the BBH model at $\theta = \pi/4$ through $E_{\text{BBH}}(\pi/4) = \frac{1}{\sqrt{2}} \left(E_{\text{SU(3)}} + 3 \right)$.

\section{iPEPS results}
\label{sec-results}

\subsection{Simple update results}

\begin{figure}[htb]
	\includegraphics[scale=0.5]{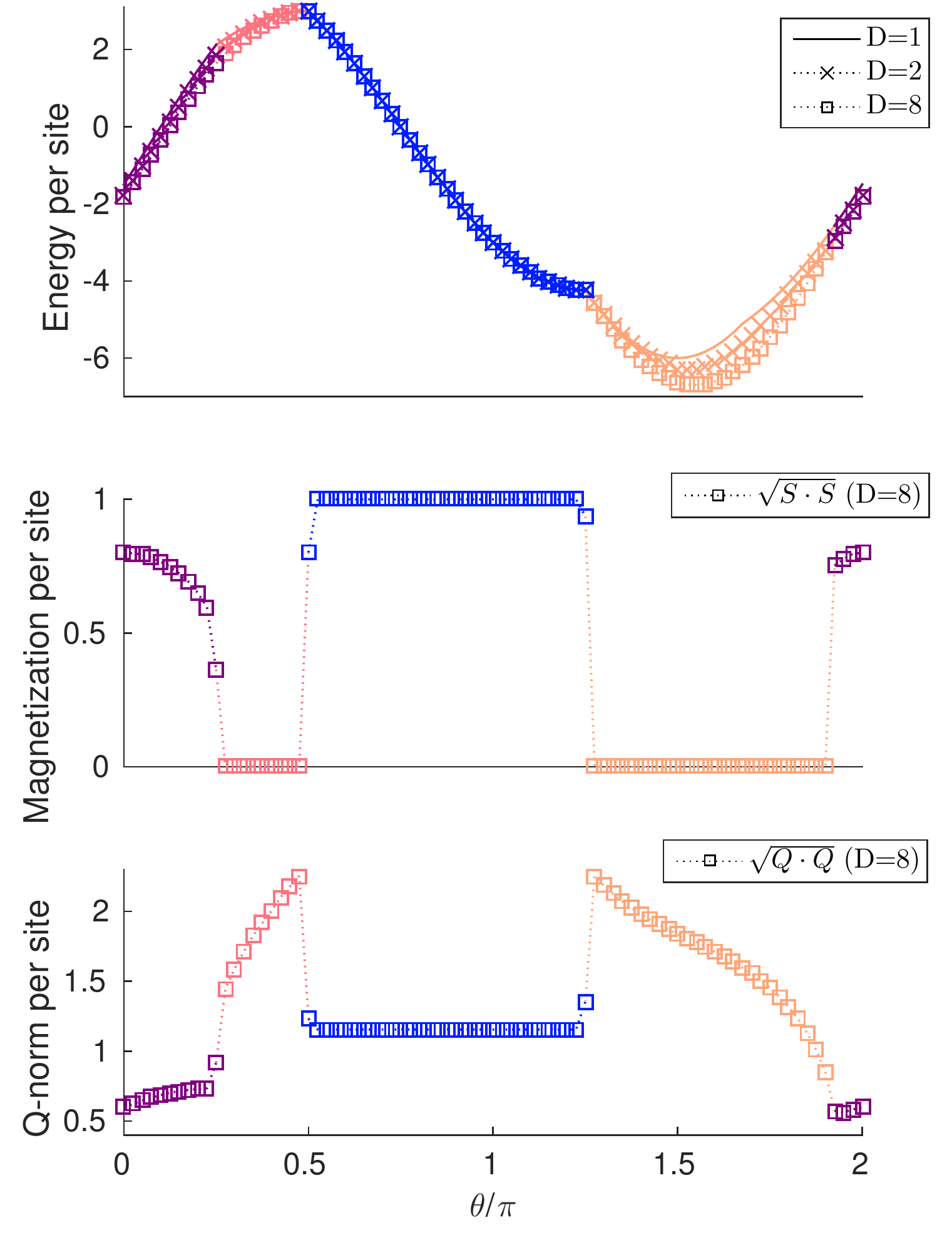}
	\caption{Energy, (\emph{top}), magnetization $m = \sqrt{\bm{S} \cdot \bm{S}}$ (\emph{middle}) and Q-norm $\sqrt{\bm{Q} \cdot \bm{Q}}$ (\emph{bottom}) per site for randomly initialized simple update simulations. From left to right we have the \textcolor{afm3Color}{AFM3},  \textcolor{afqColor}{AFQ}, \textcolor{fmColor}{FM} and  \textcolor{fqColor}{FQ} phases (color online). The magnetic phases can be recognized by a non-zero magnetization and a smaller Q-norm than that of neighboring quadrupolar phases.}
	\label{fig-simple}
	\end{figure}

	To obtain a rough picture of the phase diagram, we have performed randomly initialized simple update simulations for unit cells up to size 6x6 and bond dimensions $D=1,2,3,\ldots,8$ for 80 equidistantly spaced values of $\theta \in [0,2\pi)$. The resulting energy per site as a function of $\theta$ is shown in Fig.~\ref{fig-simple} (top). For each fixed value of $\theta$, only the lowest energy of all unit cells considered is shown.

	Observing the average magnetization and Q-norm per site displayed in the middle and bottom graphs of Fig.~\ref{fig-simple}, the simple update simulations show four different phases. Starting at $\theta = 0$, we have, in order of increasing $\theta$, the 120$\degree$ magnetically ordered (AFM3), antiferroquadrupolar (AFQ), ferromagnetic (FM) and ferroquadrupolar (FQ) phases, with transitions occurring at $\theta = \pi/4$, $\pi/2$, $5\pi/4$ and roughly $1.9\pi$ respectively. The simple update results quantitatively agree with the phase diagram found by L\"{a}uchli et al.~\cite{lauchli06}. Moreover, the jumps in the magnetization suggest that the corresponding phase transitions are of first order. 

	The simple update results do not hint at the existence of any phases other than those occurring in the product state phase diagram. However, we have learned from our study of the corresponding square lattice model~\cite{niesen17,niesen17b} that|close to transition points especially|randomly initialized simple update simulations can overlook certain phases. Therefore, we will next proceed with a more thorough full and variational update analysis, and investigate all four of the above phase transitions. In particular, we will have a look at the FQ to AFM3 transition, and provide a more accurate estimate of the critical value of $\theta$ for which the transition occurs. Note that the locations of the other three transitions are fixed, either because they sit at one of the SU(3)-symmetric points $\theta = \pi/4$ and $5\pi/4$, or because the extent ($\pi/2 < \theta < 5\pi/4$) of the FM product state phase is independent of the underlying two-dimensional lattice structure. Additionally, we will keep an eye out for a possible appearance of the $m=2/3$ phase, as well as determine the extent of the one-dimensional Haldane phase in the anisotropic triangular lattice spin-1 BBH model.

\subsection{FQ to AFM3 transition}
\label{sec-fq-afm3}

\begin{figure}[htb]
	\includegraphics[scale=0.5]{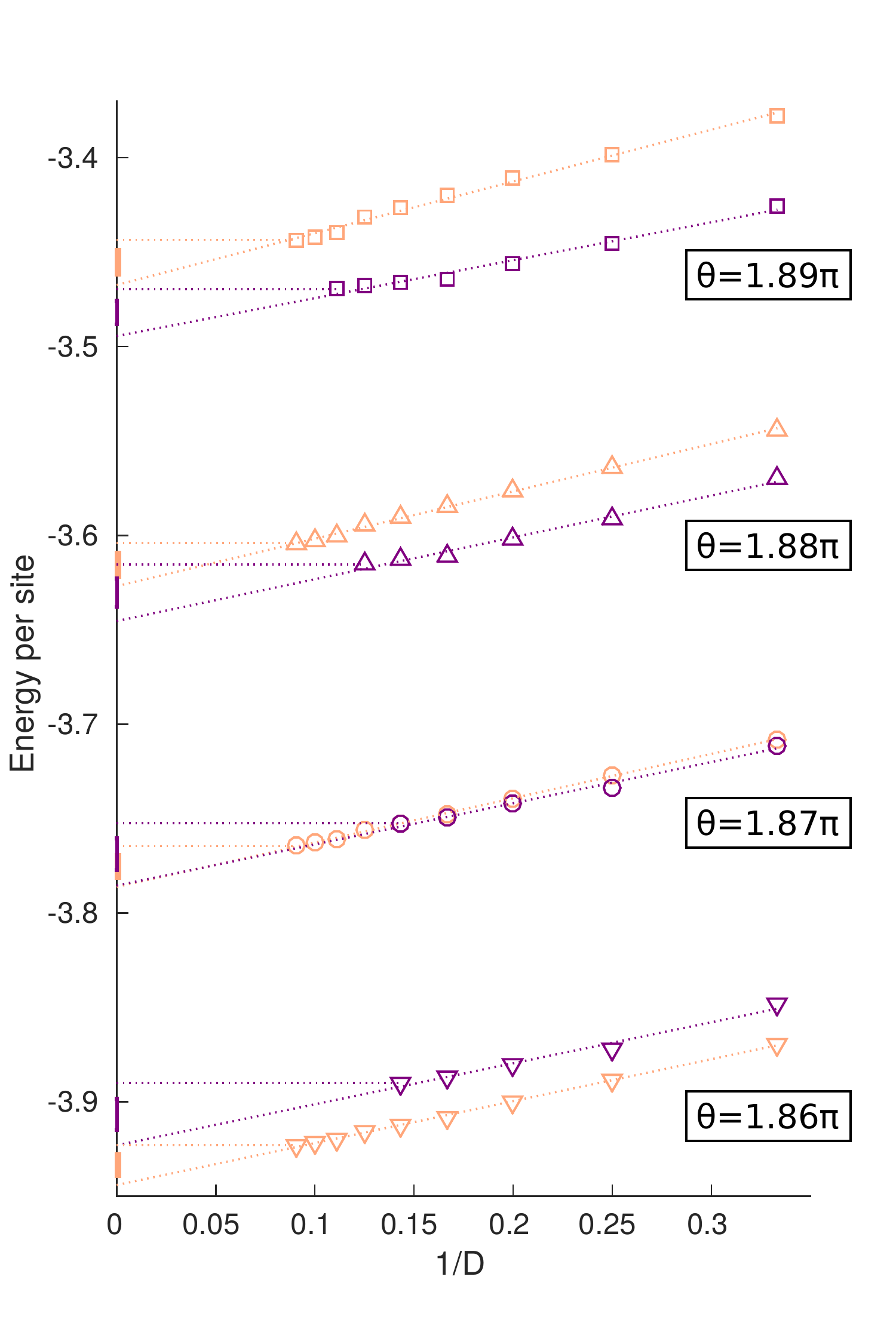}
	\caption{Energy per site (full update) for {\color{fqColor} FQ} and {\color{afm3Color} AFM3} (color online) states at $\theta = 1.86\pi - 1.89\pi$ in the vicinity of the FQ to AFM3 phase transition.}
	\label{fig-pt185-energy-D}
	\end{figure}

	Making use of hysteresis in the vicinity of a first order transition, we can simulate states in the FQ and AFM3 phases just beyond the transition point by initializing them from a state that lies deeper in the phase we want to simulate. Doing so around the simple-update-estimated transition point $\theta \approx 1.9\pi$ yields the energy per site for simulations in the FQ and AFM3 phase as shown in Fig.~\ref{fig-pt185-energy-D}. 

	For states in the FQ phase, we have imposed U(1) symmetry (aligning the on-site magnetic dipole moment along the $z$-axis), allowing us to push the full update to $D=11$.  The AFM3 states, however, break U(1) symmetry because the spins do not align along a given axis, and we can therefore go up to $D=9$ at best~\cite{n-datapoints}. 

\begin{figure}[htb]
	\includegraphics[scale=0.5]{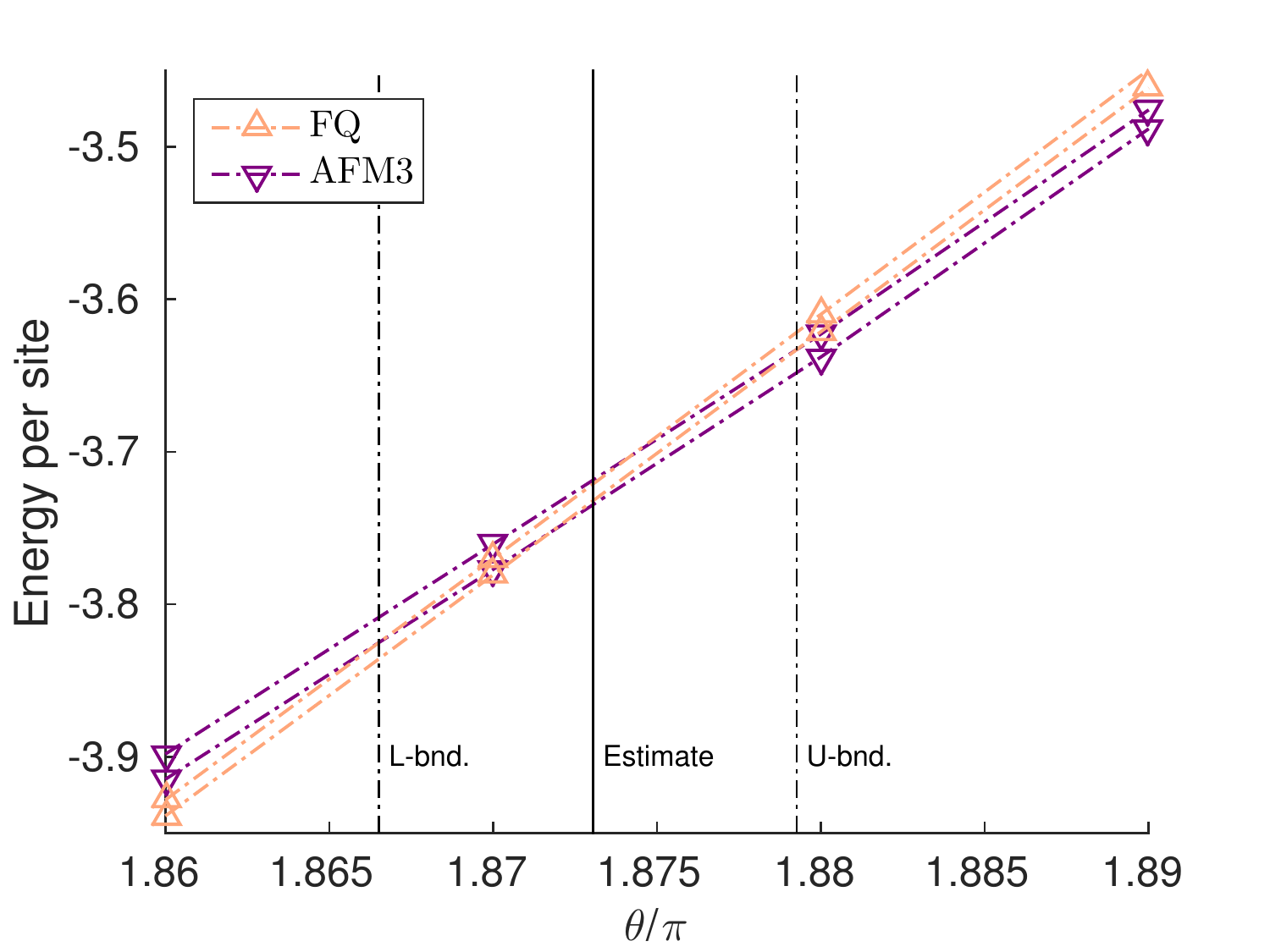}
	\includegraphics[scale=0.5]{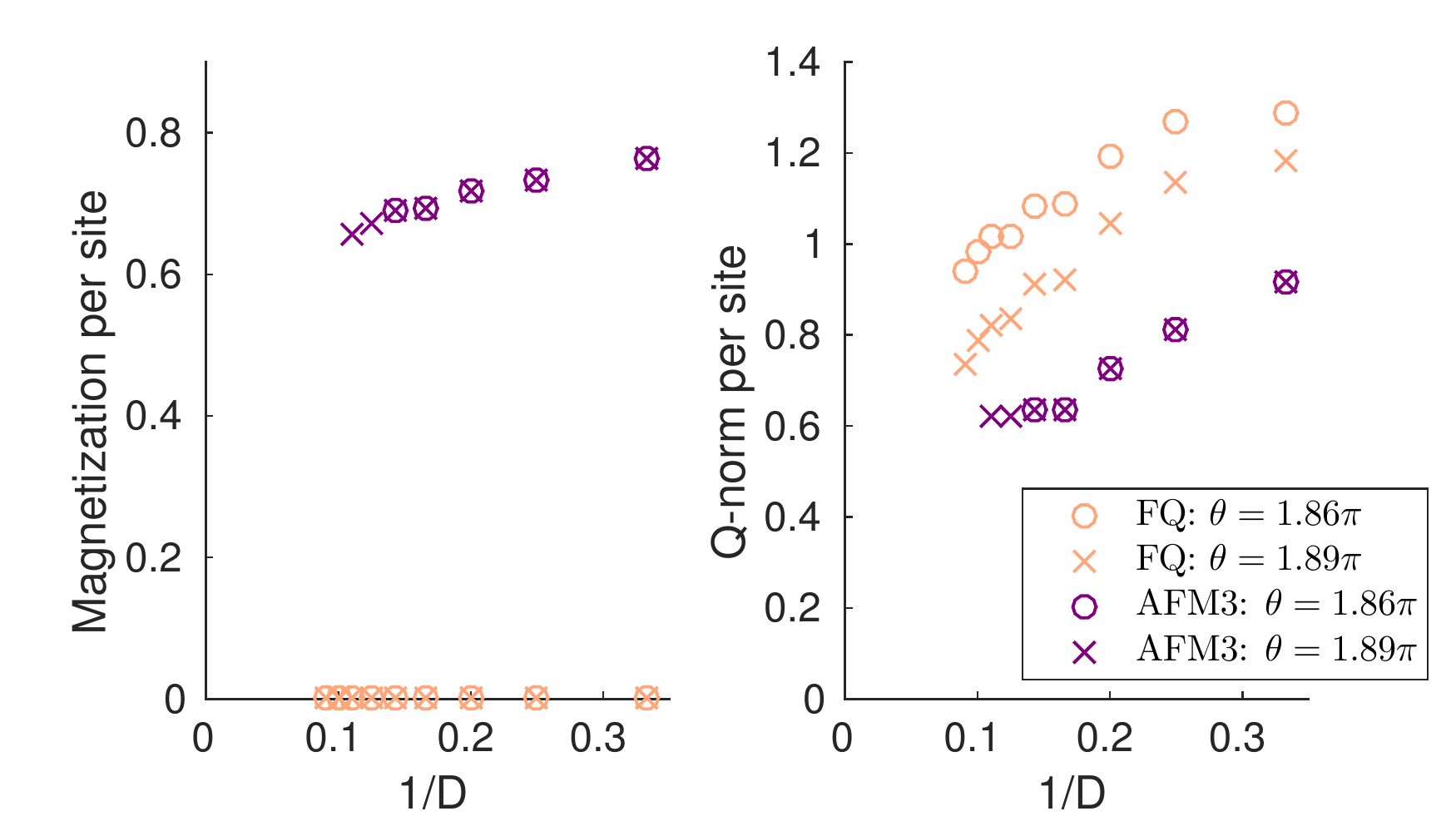}
	\caption{\emph{Top}: extrapolated energy per site (top and bottom of error bars from Fig.~\ref{fig-pt185-energy-D} shown) for the FQ and AFM3 states around the FQ to AFM3 phase transition. The vertical lines signified with ``L-bnd." and ``U-bnd." indicate the locations for which the error bars separate. The ``Estimate" is the intersection point of the curves drawn through the centers of the error bars shown in Fig.~\ref{fig-pt185-energy-D}. We conclude that the phase transition occurs at $\theta_c = 1.873(7)\pi$. \emph{Bottom}: Magnetization and Q-norm per site for FQ and AFM3 states left ($\theta = 1.86\pi$) and right ($\theta = 1.89\pi$) of the phase transition. The magnetization especially displays a clear jump when going from the FQ to AFM3 phase, which, combined with the slight kink in the energy demonstrates that this transition is first order.}
	\label{fig-pt185}
	\end{figure}

	Extrapolating $D\rightarrow \infty$ as explained in Section~\ref{sec-benchmk} yields an estimated energy per site and corresponding error bar for both the FQ and AFM3 simulations at different values of $\theta$. Plotting the upper and lower bounds of the error bars as a function of $\theta$ (top Fig.~\ref{fig-pt185}) then gives an estimate for the critical value of $\theta$ that separates the FQ and AFM3 phases of $\theta_c = 1.873(7)\pi$. This result is a more accurate refinement of the exact diagonalization result extrapolated to infinite system size ($\theta_c^{ED} \approx 1.89\pi$) obtained by L\"{a}uchli et al.~\cite{lauchli06}.

	The fact that the energy per site curves for the FQ and AFM3 simulations have (slightly) different slopes in Fig.~\ref{fig-pt185} implies that the energy per site of the ground state displays (a slight) kink at the FQ to AFM3 intersection. Supplemented by the jump in magnetization and the different $1/D$ behavior of the quadrupole norm|displayed in the bottom plots of Fig.~\ref{fig-pt185}|we can conclude that the FQ to AFM3 transition is first order.

\subsection{AFQ to FM transition and absence of m=2/3 phase}

\begin{figure}[htb]
	\includegraphics[scale=2]{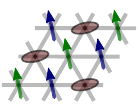}
	\caption{The three-sublattice $m=2/3$ state with ferromagnetically aligned magnetic moments on two sublattices and a director|i.e. a normal vector to the plane of fluctuations|on the third sublattice that is parallel to the neighboring magnetic moments.}
	\label{fig-m23}
	\end{figure}

	The $m=2/3$ magnetization plateau in the AFQ phase found by L\"{a}uchli et al.~\cite{lauchli06} at finite magnetic field (mentioned in Section~\ref{sec-prevst}) corresponds to a three-sublattice state with magnetic moments ferromagnetically aligned on two of the sublattices, and on the third a quadrupolar director parallel to the magnetic moments on the neighboring sites (Fig.~\ref{fig-m23}). L\"{a}uchli et al.~discovered that, as $\theta$ increases towards $\pi/2$, the value of the external magnetic field for which the transition to the $m=2/3$ phase occurs decreases as $\theta$ increases, up to the critical point $\theta = \pi/2$ where the AFQ and $m=2/3$ states are simultaneous ground states of the zero-external-field BBH model. On the square lattice, T\'{o}th et al.~\cite{toth12} showed that a very similar phenomenon occurs (in that case, the partially-magnetized state was half-magnetized instead of two-thirds). Thus, in light of our recent discovery~\cite{niesen17b} of the half-magnetized phase actually taking up a non-negligible portion of the square lattice zero-field phase diagram, it seems natural to ask whether the $m=2/3$ phase also occurs on the triangular lattice BBH model with zero external field.

\begin{figure}[htb]
	\includegraphics[scale=0.5]{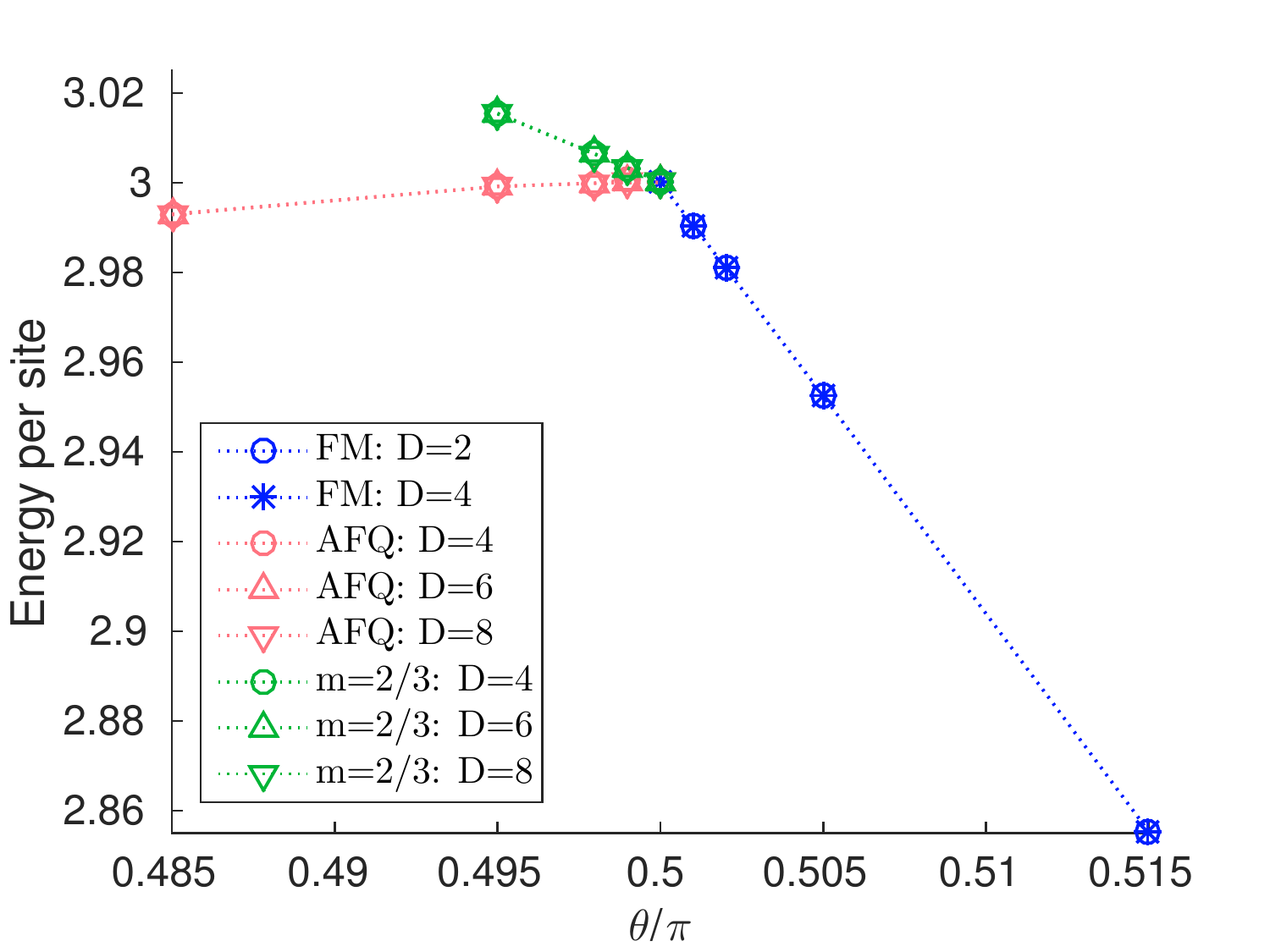}
	\hspace*{-0.5cm} \includegraphics[scale=0.5]{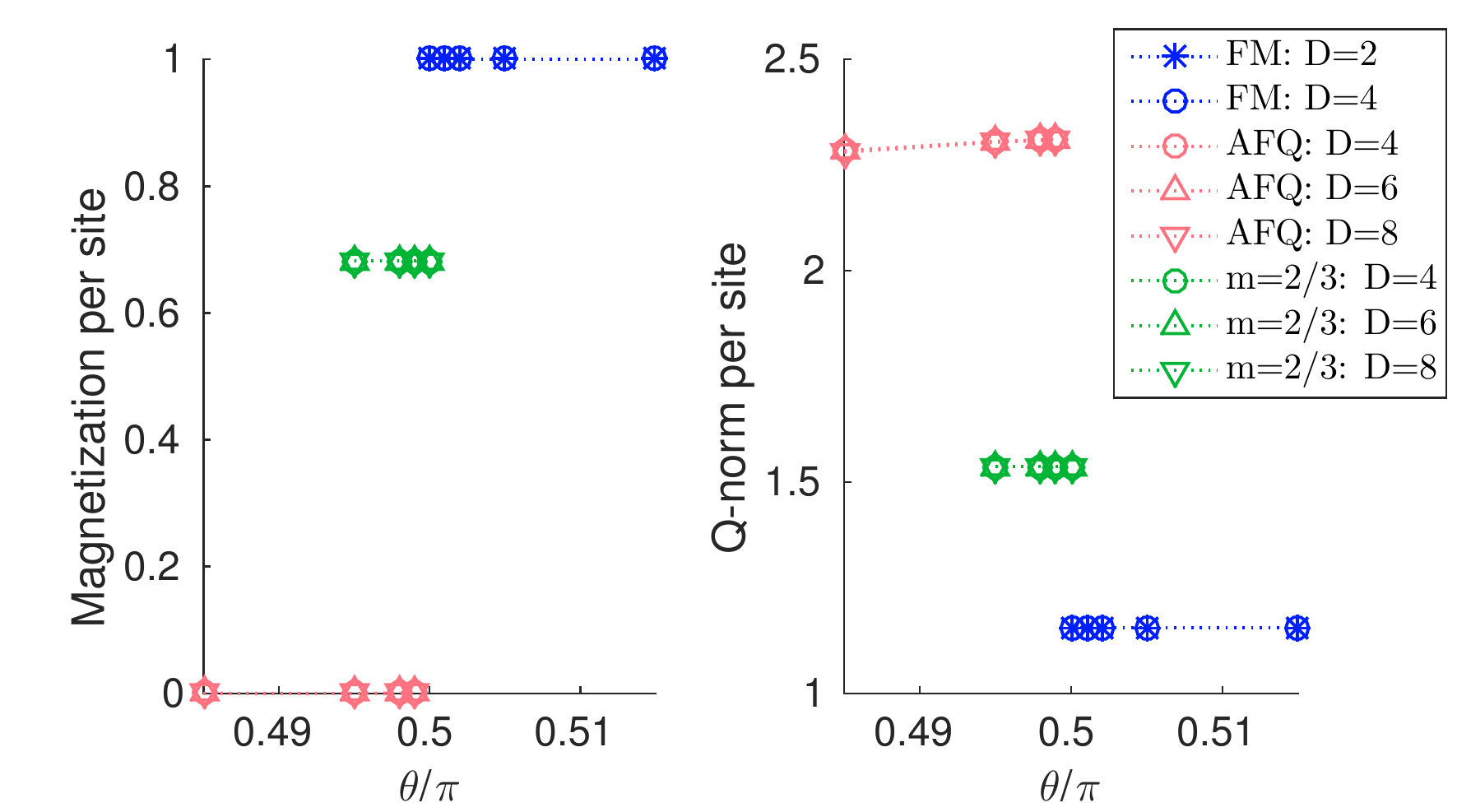} 
	\caption{\emph{Top}: energy per site (full update) for AFQ, FM and $m=2/3$ states around $\theta = \pi/2$. Because all states are (practically) product states, the energies do not depend on $D$. \emph{Bottom}: Magnetization and Q-norm per site (full update) for AFQ, FM and $m=2/3$ states around $\theta = \pi/2$. The jumps in both and kink in the energy show that the transition is first order.}
	\label{fig-pt050}
	\end{figure}

	We initialized several simulations in the vicinity of $\theta = \pi/2$. The energy per site of the AFQ, FM and $m=2/3$ simulations is shown in Fig.~\ref{fig-pt050}. We can conclude that, contrary to the square lattice case, the $m=2/3$ states are everywhere higher in energy than the AFQ states|except at the AFQ to FM transition point where the ground state is degenerate and the $m=2/3$ state is one of the many ground states|and thus the $m=2/3$ phase does not occur in the zero-field phase diagram. 

	Note that, in the FM phase the ground state is a product state. In the vicinity of the FM phase, the ground state is very close to a product state, as can be seen from the fact that the energy does not visibly improve with increasing bond dimension. Therefore, we do not have to do $D \rightarrow \infty$ extrapolations to get accurate results. 

	From the clear kink in the energy per site, and the jumps in magnetization and Q-norm (Fig.~\ref{fig-pt050}), we can conclude that the AFQ to FM transition is also of first order.

	Finally, let us have a look at the remaining two phase transitions, located at the SU(3)-symmetric points $\theta = \pi/4$ and $\theta = 5\pi/4$. The results we find agree with previous studies, and will be presented for completeness.

\subsection{AFM3 to AFQ transition}

\begin{figure}[htb]
	\includegraphics[scale=0.5]{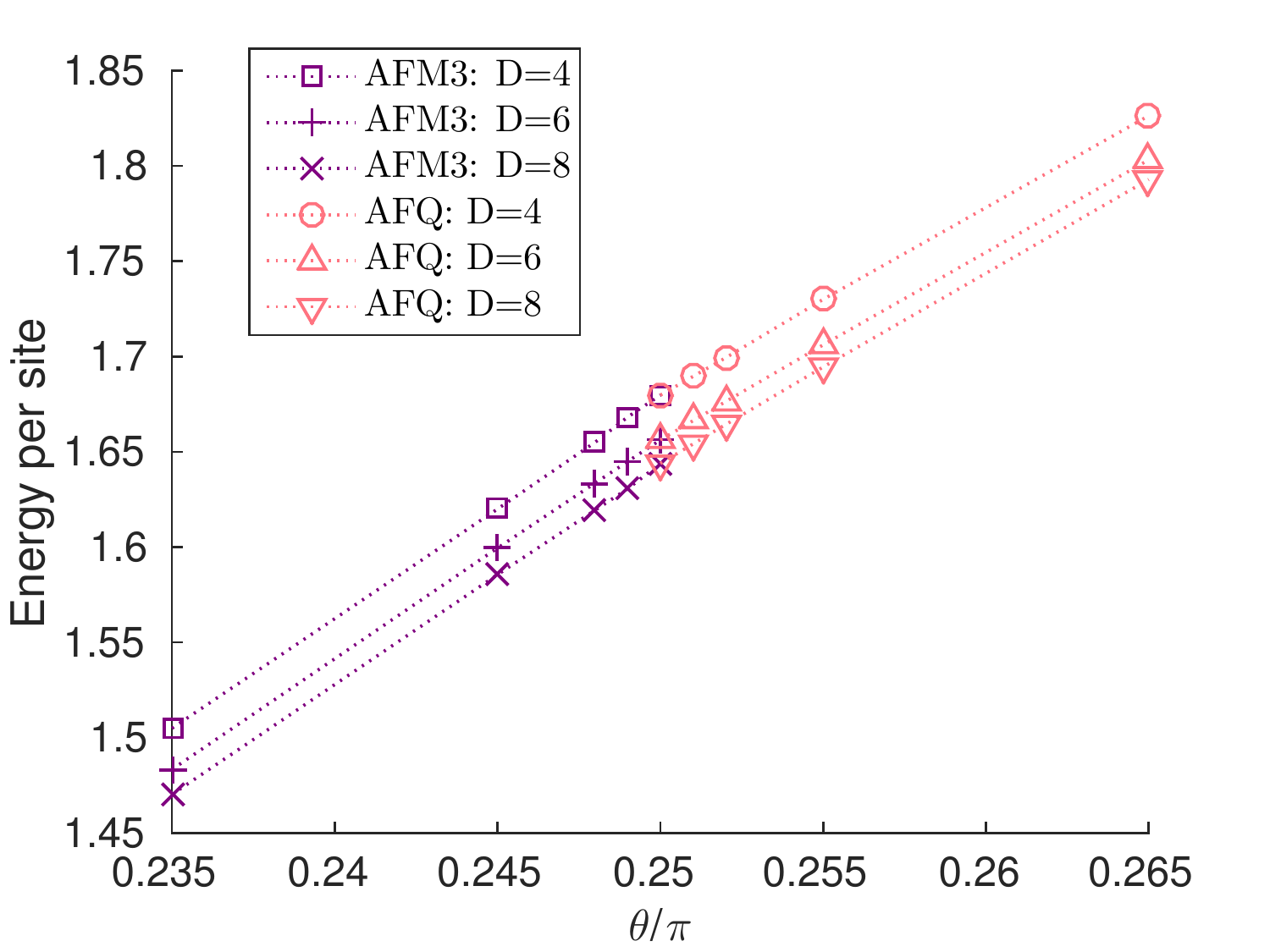}
	\includegraphics[scale=0.5]{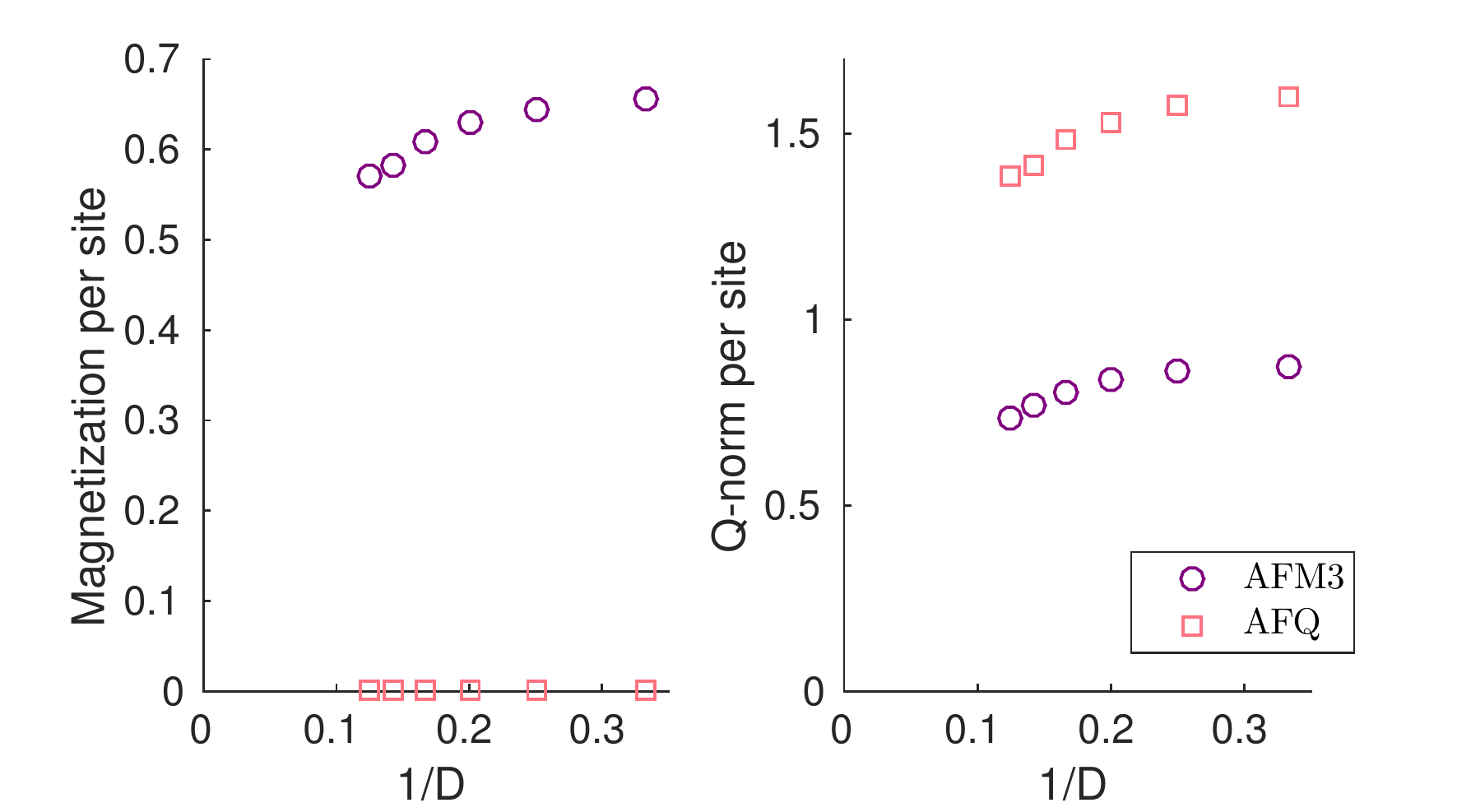}
	\caption{\emph{Top}: energy per site (full update) for $D=4$, $6$ and $8$ for AFM3 and AFQ states around $\theta = \pi/4$. \emph{Bottom}: Magnetization and Q-norm per site (full update) for AFM3 and AFQ states exactly at $\theta = \pi/4$. The jumps in both and slight kink in the energy for fixed values of $D$ show that the AFM3 to AFQ transition is of first order.}
	\label{fig-th025}
	\end{figure}

	Approaching the phase transition at the SU(3) point $\theta = \pi/4$ from both the AFM3 and AFQ sides by slowly walking towards the critical point, loading each simulation from the last (for fixed $D$), we obtain the energy per site plot shown in Fig.~\ref{fig-th025}.

	Moving towards the transition as described above, we can ensure that simulations stay in their respective phases even at the critical point itself (where both $D \rightarrow \infty$ extrapolated AFM3 and AFQ states are ground states of the system). Fig.~\ref{fig-th025} shows the resulting magnetization and Q-norm exactly at the transition at $\theta = \pi/4$. The subtle kink in the fixed-$D$ energy per site plots and the jumps in magnetization and Q-norm show that the transition is first order.

\subsection{FM to FQ transition}

	The FM to FQ phase transition can be investigated in the same manner as the AFQ to FM transition. As noted by V\"{o}ll et al.~\cite{voll15}, the ground state in (and close to) the FM phase is a product (or almost product) state, implying that no $D\rightarrow \infty$ extrapolation will be required. 

	\begin{figure}[htb]
	\includegraphics[scale=0.5]{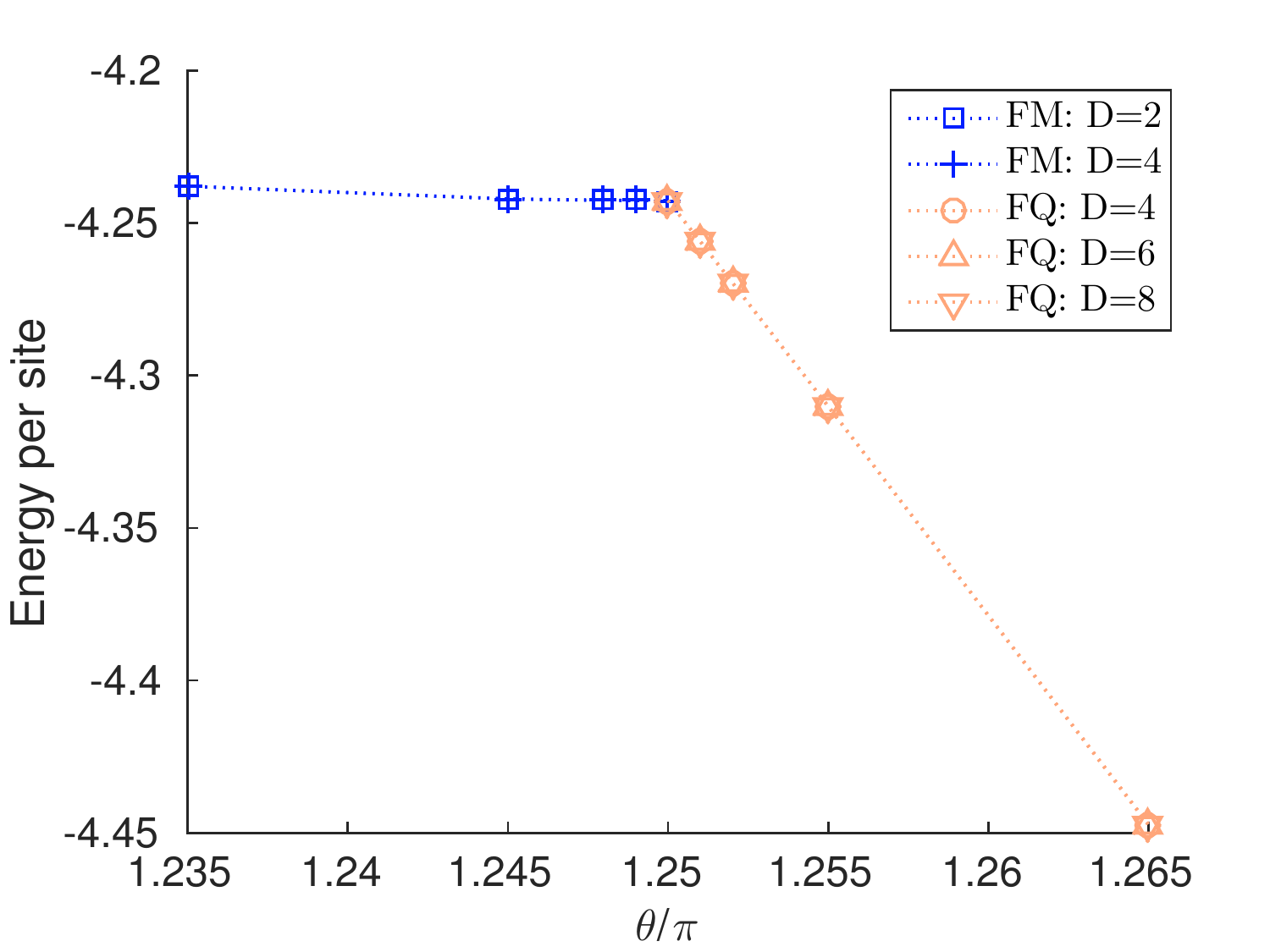}
	\includegraphics[scale=0.5]{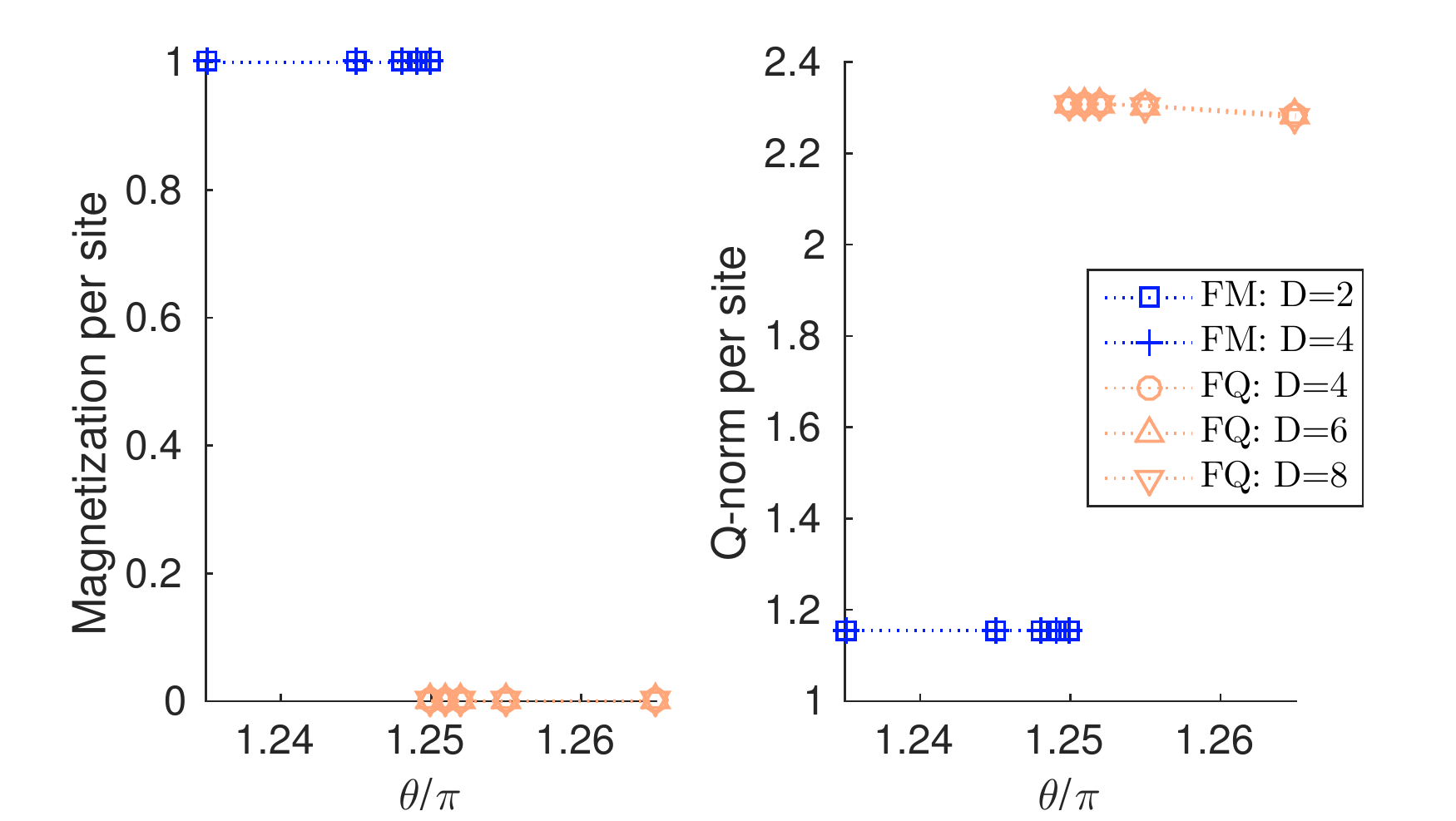}
	\caption{\emph{Top}: energy per site (full update) for FM and FQ states around $\theta = 5\pi/4$. The FM states are product states, and the FQ states are close to being product states as can be seen from the fact that the energies do not improve with $D$. \emph{Bottom}: Magnetization and Q-norm per site (full update) for FM and FQ states around $\theta = 5\pi/4$. The jumps in magnetization and Q-norm and the kink in the energy support the claim that this transition is first order.}
	\label{fig-th125}
	\end{figure}

	The clear kink in the energy per site and jumps in magnetization and Q-norm (Fig.~\ref{fig-th125}) show that the FM to FQ phase transition is also of first order.

\subsection{Haldane phase in the anisotropic model}

	Our previous investigation~\cite{niesen17} of the square lattice BBH model showed that in between the ordinary antiferromagnetic and the 120$\degree$ magnetically ordered phases a quantum paramagnetic phase arises that can be adiabatically connected to the Haldane phase of decoupled one-dimensional spin-1 BBH chains. On the triangular lattice, there is no competition between two and three-sublattice order. Nevertheless, the fact that the FQ to AFM3 product ground state phase transition point at $\theta = \arctan(-2) \approx 1.658\pi$ shifts significantly to $\theta = 1.873(7)\pi$ shows that quantum fluctuations play an important role in the FQ to AFM3 transition, and perhaps also allow for the possibility of an intermediate quantum paramagnetic phase.

	The first sign that hinted at the presence of a quantum paramagnetic phase on the square lattice was the vanishing magnetization in the antiferromagnetic phase. Looking at Fig.~\ref{fig-pt185}, the magnetization in the AFM3 phase clearly does not vanish in the $D \rightarrow \infty$ limit, but we do observe that the quadrupolar order of the FQ simulation at $\theta = 1.89\pi$ goes down as $D$ increases. However, it would be too strong a claim to say that it extrapolates to zero. Besides, $\theta = 1.89\pi$ is already in the AFM3 phase, as the extrapolated FQ energy is higher than the extrapolated AFM3 energy. Thus, based on the full update results in Fig.~\ref{fig-pt185}, there is no intermediate paramagnetic phase in between the FQ and AFM3 phases.

	It is possible that the FQ to AFM3 transition is not the right place to look for a paramagnetic ground state. Motivated by the emergence of the one-dimensional Haldane phase on the two-dimensional square lattice, a natural starting point for looking for a quantum paramagnetic phase is to investigate the extent of the Haldane phase on the anisotropic triangular lattice. Because the one-dimensional spin-1 BBH chain lies in the Haldane phase for $-\pi/4 < \theta < \pi/4$, this is the parameter range will shall focus on.

	We introduce an additional coupling parameter $0 \leq J_{\text{anis}} \leq 1$ that modifies the diagonal and vertical bonds of the triangular lattice simultaneously; $J_{\text{anis}} = 0$ corresponding to the limit of decoupled horizontal one-dimensional chains, and $J_{\text{anis}} = 1$ corresponding to the isotropic two-dimensional triangular lattice.
	
\begin{figure}[htb]
	\includegraphics[scale=0.55]{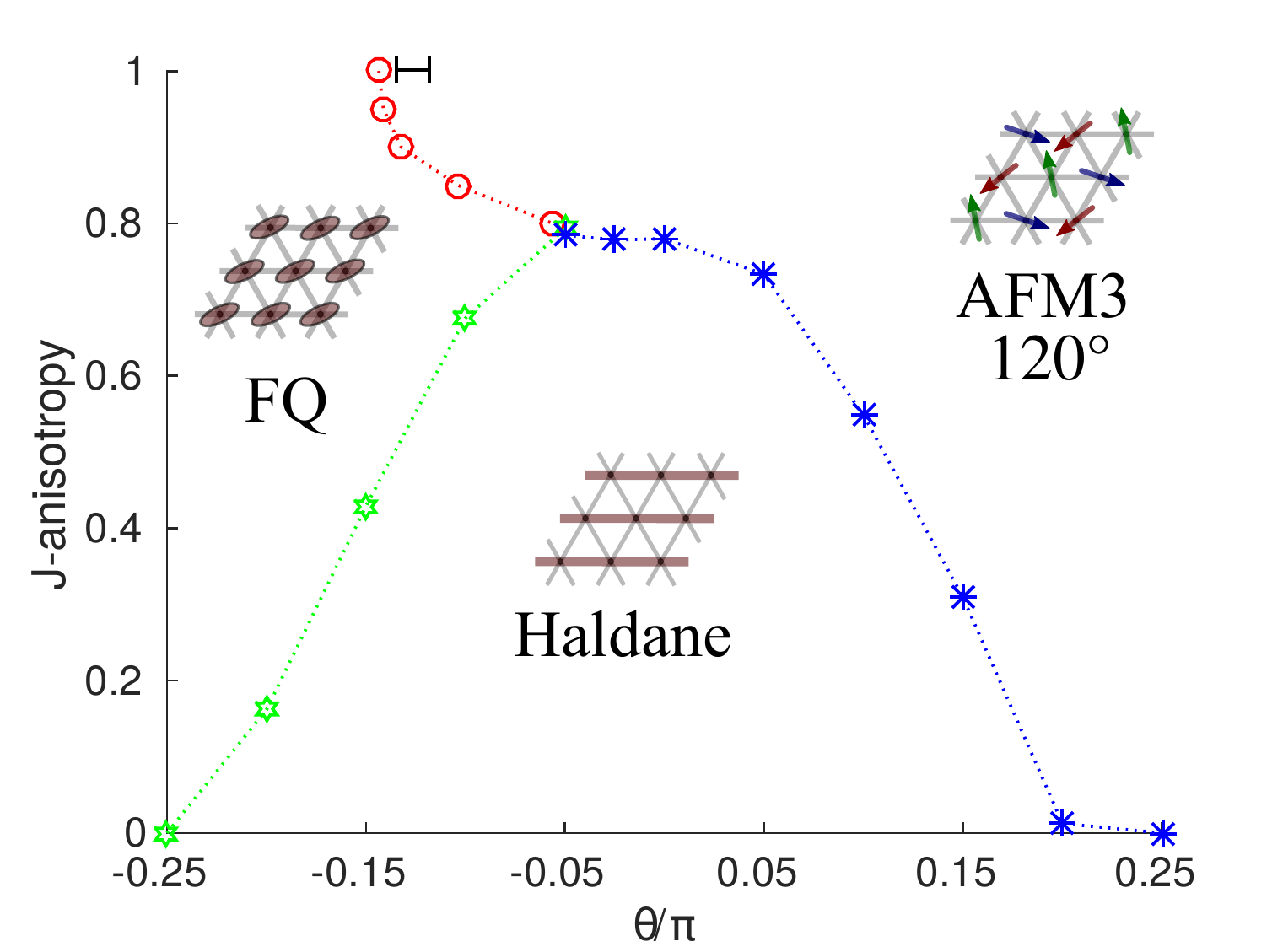}
	\caption{$D=9$ simple update phase diagram of the \mbox{spin-1} BBH model on the anisotropic triangular lattice. The full update result $\theta_c = 1.873(7)\pi$ for the location of the FQ to AFM3 phase transition at $J_{\text{anis}} = 1$ indicated by the black error bar is shown to give an idea of the accuracy of the $D=9$ simple update phase diagram.}
	\label{fig-anis-pd}
	\end{figure}

	To map the entire $\theta$-$J_{\text{anis}}$ phase diagram using full updates and $D \rightarrow \infty$ extrapolation is computationally too expensive. Thus, we shall revert to a fixed $D=9$ simple update investigation. The result is plotted in Fig.~\ref{fig-anis-pd}. Note that we also looked for additional phases other than the FQ, AFM3 and Haldane phases|by running simulations with randomly initialized tensors scattered throughout the $\theta$-$J_{\text{anis}}$ plane|but we did not encounter other types of order.

	To estimate of the accuracy of the simple update result, we can use the FQ to AFM3 transition point at $J_{\text{anis}} = 1$ computed in Section~\ref{sec-fq-afm3}. Because the critical $\theta_c^{\text{simple}}$ separating the FQ and AFM3 phases predicted by the fixed $D=9$ simple update lies just outside the error bar of the full update result $\theta_c = 1.873(7)\pi$, we can expect that the true phase separation lines lie in the vicinity of those shown in Fig.~\ref{fig-anis-pd}, but their precise location cannot be inferred form the plot. However, the $D=9$ simple update phase diagram does seem accurate enough to conclude that the Haldane phase does \emph{not} extend all the way up to the isotropic limit. 

	From Fig.~\ref{fig-anis-pd} we observe that the Haldane phase extends maximally in the vicinity of the Heisenberg point $\theta = 0$ (where the biquadratic coupling is zero) rather than at the FQ to AFM3 transition point. As an extra check, we have pushed our simulations at the Heisenberg point to high $D$ (Appendix~\ref{app-add-data} Fig.~\ref{fig-th0}) to verify that the magnetization stays non-zero in the $D\rightarrow \infty$ limit. In addition, we have also done full update simulations for $J_{\text{anis}} = 1$ initialized directly from within the Haldane phase and compared the energy to that of the FQ and AFM3 simulations (both close to the FQ to AFM3 transition|Appendix~\ref{app-add-data} Fig.~\ref{fig-th186-hald}, and at the Heisenberg point|Appendix~\ref{app-add-data} Fig.~\ref{fig-th0-hald}); the result of which shows that clearly, the Haldane simulations are much higher in energy in the isotropic model.

	We note that the Haldane phase in the extreme anisotropic limit might a priori be better approximated by an anisotropic iPEPS (with a larger bond dimension in the $x$-direction). However, by comparing the weights on the $x$ and $y$-bonds obtained in the simple update approach~\cite{jiang08} in the Haldane phase at large $D$, we observe that, as $J_{\text{anis}}$ increases, the smallest weights on the $x$ and $y$-bonds become of similar magnitude~\cite{n-D9bonds}, showing that an isotropic ansatz is appropriate here. (An isotropic ansatz was also used in Ref.~\cite{niesen17} to determine the phase boundary between the Haldane and antiferromagnetic phases on the square lattice spin-1 BBH model, but even in the strongly anisotropic limit, the iPEPS result was found to be very close to the reference value from Quantum Monte Carlo.)

	Combining our results from the simple update anisotropic phase diagram, the full update study at the Heisenberg point and the full update Haldane simulations at the isotropic limit, we can safely conclude that the Haldane phase does \emph{not} extend all the way up to the isotropic triangular lattice $J_{\text{anis}} = 1$. Moreover, we did not encounter any other signs that hint at the presence of a quantum paramagnetic phase, and we can therefore conclude that the phase diagram as shown in Fig.~\ref{fig-ipeps-pd} is the complete ground state phase diagram.

\section{Conclusion}
\label{sec-conclusion}

	We have presented a complete and systematic iPEPS study of the ground state phase diagram of the spin-1 bilinear-biquadratic Heisenberg (BBH) model on the triangular lattice. We found the ferromagnetic and 120$\degree$ magnetically ordered as well as the ferro and antiferroquadrupolar phases, and precisely determined that the ferroquadrupolar to 120$\degree$ magnetically ordered phase transition occurs at $\theta_c = 1.873(7)\pi$. This number is close to the exact diagonalization estimate by L\"{a}uchli et al.~\cite{lauchli06} that predicted the transition to occur at $\theta_c^{\text{ED}} \approx 1.89\pi$. Moreover, our simulations show that the partially-magnetic partially-quadrupolar phase that we encountered on the square lattice~\cite{niesen17b} does not appear on the triangular lattice spin-1 BBH model (Fig.~\ref{fig-pt050}).

	Inspired by our finding~\cite{niesen17} of the one-dimensional Haldane phase extending all the way to the two-dimensional isotropic limit on the square lattice, we searched for signs of a possible quantum paramagnetic phase on the triangular lattice. The simple update results (Fig.~\ref{fig-simple}) did not hint at the presence of a quantum paramagnetic phase. At the ferroquadrupolar to 120$\degree$ magnetically ordered phase transition, we observed that the Q-norm of the ferroquadrupolar simulations decreases as we approach the magnetic phase (Fig.~\ref{fig-pt185}), but it still extrapolates to a non-zero number even beyond the phase transition (at $\theta = 1.89\pi$) in the magnetically ordered phase.

	We then investigated the extent of the Haldane phase on the anisotropic triangular lattice (Fig.~\ref{fig-anis-pd}) and found that, close to the Heisenberg point $\theta = 0$, it extends maximally to approximately $J_{\text{anis}} \approx 0.8$. However, both the high $D$ full update simulations at the Heisenberg point as well as the full update simulations initialized directly from within the Haldane phase (Appendix~\ref{app-add-data}) confirm that the ground state is ordered, reaffirming that the ground state phase diagram of the triangular lattice \mbox{spin-1} BBH model is as depicted in Fig.~\ref{fig-ipeps-pd}, and in particular does not contain a quantum paramagnetic phase. Surprisingly then, for the spin-1 BBH model, quantum effects seem to have less surprising consequences on the (in the AFM phase) geometrically frustrated triangular lattice than they do on the square lattice. 

	Because the triangular lattice has a larger coordination number than the square lattice, it seems tempting to hypothesize that the Haldane phase is unlikely to occur for densely connected lattices. Intuitively, this makes sense, because the energy gained by forming valence bonds in one particular direction at the cost of increased energy on the remaining bonds seems beneficial only when there are not too many remaining bonds. For lattices with a small coordination number on the other hand, if a possibility exists to form short valence bond loops|which is the case on the honeycomb~\cite{zhao12} and Kagome~\cite{changlani15} lattices| the ground state will break translational symmetry by forming loops of length six and three respectively. Also, a very recent study of the spin-1 BBH model on the star-shaped lattice by Lee and Kawashima~\cite{lee18} shows that a spin-liquid-like phase appears (in a region that encompasses the region in which we found the Haldane phase on the square lattice) that a priori does not seem to be connected to a one-dimensional state. Only on the square lattice the system prefers infinitely-long Haldane chains over four-site valence bond loops.

	Our study of the anisotropic model does reveal that spin-1 materials with a triangular lattice structure that are effectively described by a simple Heisenberg antiferromagnetic coupling ($\theta = 0$) are quite sensitive to anisotropies. Indeed, based on our simple update calculations we expect a transition to the Haldane phase at around $J_{\text{anis}}^{\text{triang}} \approx 0.8$, which is significantly larger than the value of $J_{\text{anis}}^{\text{square}} \approx 0.04$ for which the same transition occurs at the Heisenberg point on the square lattice. This fact could possibly be used for future experimental research that attempts to realize the extended Haldane phase in an actual two-dimensional material.

	From the perspective of tensor network methods|viewing the triangular lattice as a square lattice with additional diagonal next-nearest neighbor interactions|we would like to point out that this is one of the few systematic full update studies of models beyond nearest-neighbor interactions (see also Refs.~\cite{corboz13,corboz14b,poilblanc17,haghshenas18b,chen18}).

	Lastly, having accurately established the ground state phase diagram of the triangular lattice spin-1 BBH model in the thermodynamic limit by means of an unbiased method, future research can more confidently look at exited states or additions to the Hamiltonian beyond the biquadratic interaction in search for an explanation of the unusual behavior~\cite{tsunetsugu06,tsunetsugu07,li07,bhattacharjee06,stoudenmire09,nakatsuji05,nakatsuji07,nakatsuji10} of NiGa$_2$S$_4$ and the 6$H$-$B$ phase~\cite{cheng11,fak17,xu12,serbyn11,bieri12} of Ba$_3$NiSb$_2$O$_9$,

\acknowledgements

	We acknowledge useful discussions on the AFM to Haldane transition with K. Totsuka, O. Starykh, and S. Sachdev. This project has received funding from the European Research Council (ERC) under the European Union's Horizon 2020 research and innovation programme (grant agreement No 677061). This work is part of the \mbox{Delta-ITP} consortium, a program of the Netherlands Organization for Scientific Research (NWO) that is funded by the Dutch Ministry of Education, Culture and Science~(OCW).

\appendix

\section{Additional data on the Haldane-initialized simulations}
\label{app-add-data}

\subsection{The Heisenberg point}

\begin{figure}[htb]
	\includegraphics[scale=0.5]{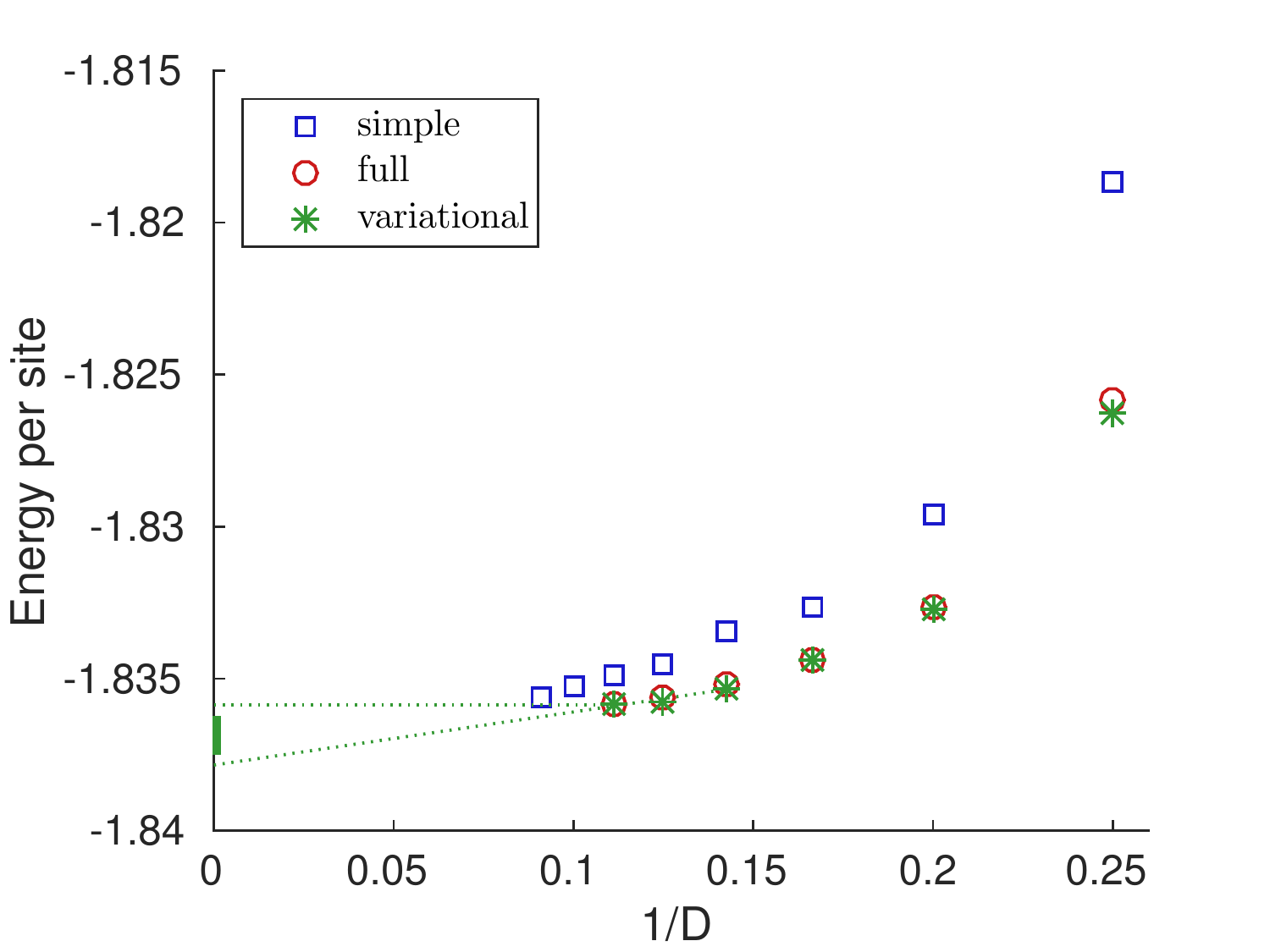}
	\includegraphics[scale=0.5]{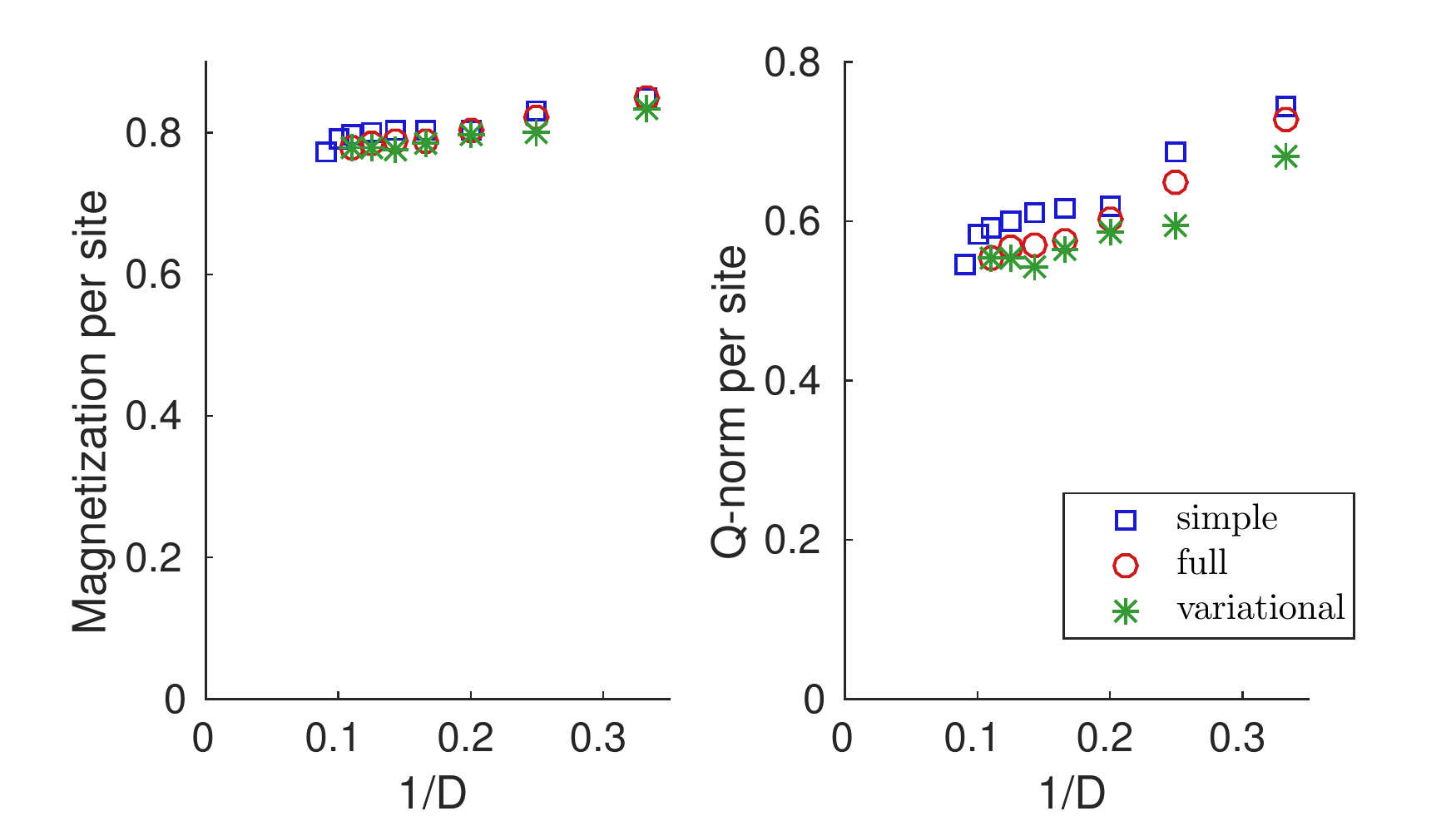}
	\caption{The energy (\emph{top}) and the magnetization and Q-norm (\emph{bottom}) per site for the simple, full and variational update algorithm at the Heisenberg point $\theta = 0$. The variational update results extrapolate to a ground state energy of $E^{\text{var}}_{D\rightarrow \infty} =  -1.8368(5)$.}
	\label{fig-th0}
	\end{figure}

	We have looked for signs of a vanishing magnetization at the Heisenberg point ($\theta = 0$)|a point that lies in the region where the Haldane phase extends furthest in the $\theta$-$J_{\text{anis}}$ phase diagram towards the isotropic limit (Fig.~\ref{fig-anis-pd})|by pushing the simple update to $D=11$, and the full and variational update to $D=9$. The resulting energy, magnetization and Q-norm per site are plotted in Fig.~\ref{fig-th0}.

	The full and variational update simulations (Fig.~\ref{fig-th0}) clearly extrapolate to a finite magnetic and quadrupole moment, and we can therefore conclude that the ground state is ordered at $\theta = 0$. The less-accurate simple update magnetization and Q-norm seem to curve downwards, but this is likely an artifact of the simple update, as the $D=11$ simple update result is very similar in energy, magnetization and quadrupole moment to the $D=8$ full and variational update results. Moreover, magnetization and quadrupole curves as a function of $1/D$ typically do not lie on a perfectly straight line, so not too much importance should be given to an individual data point.

\subsection{Haldane simulations at the isotropic limit}

\begin{figure}[htb]
	\includegraphics[scale=0.5]{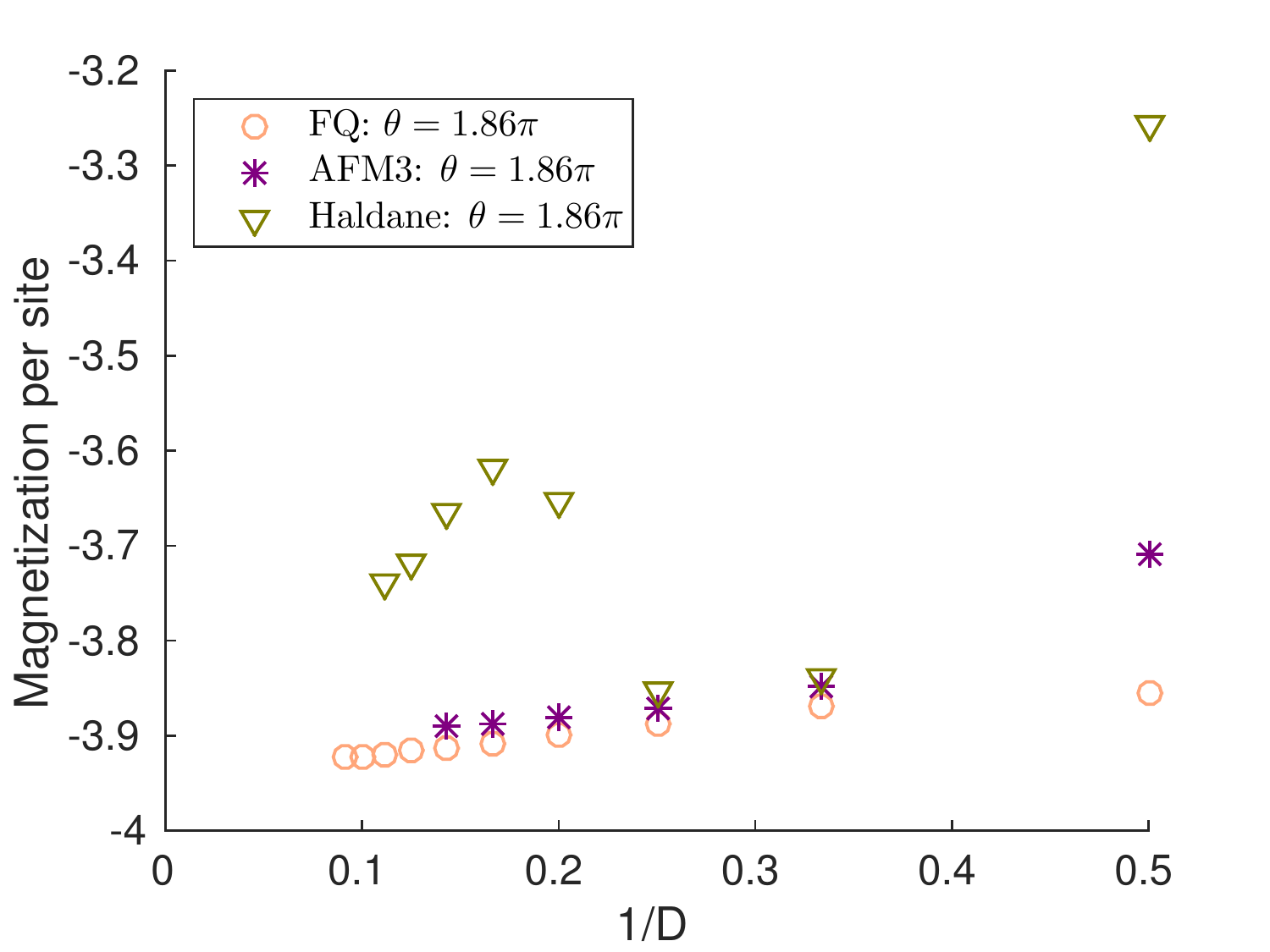}
	\includegraphics[scale=0.5]{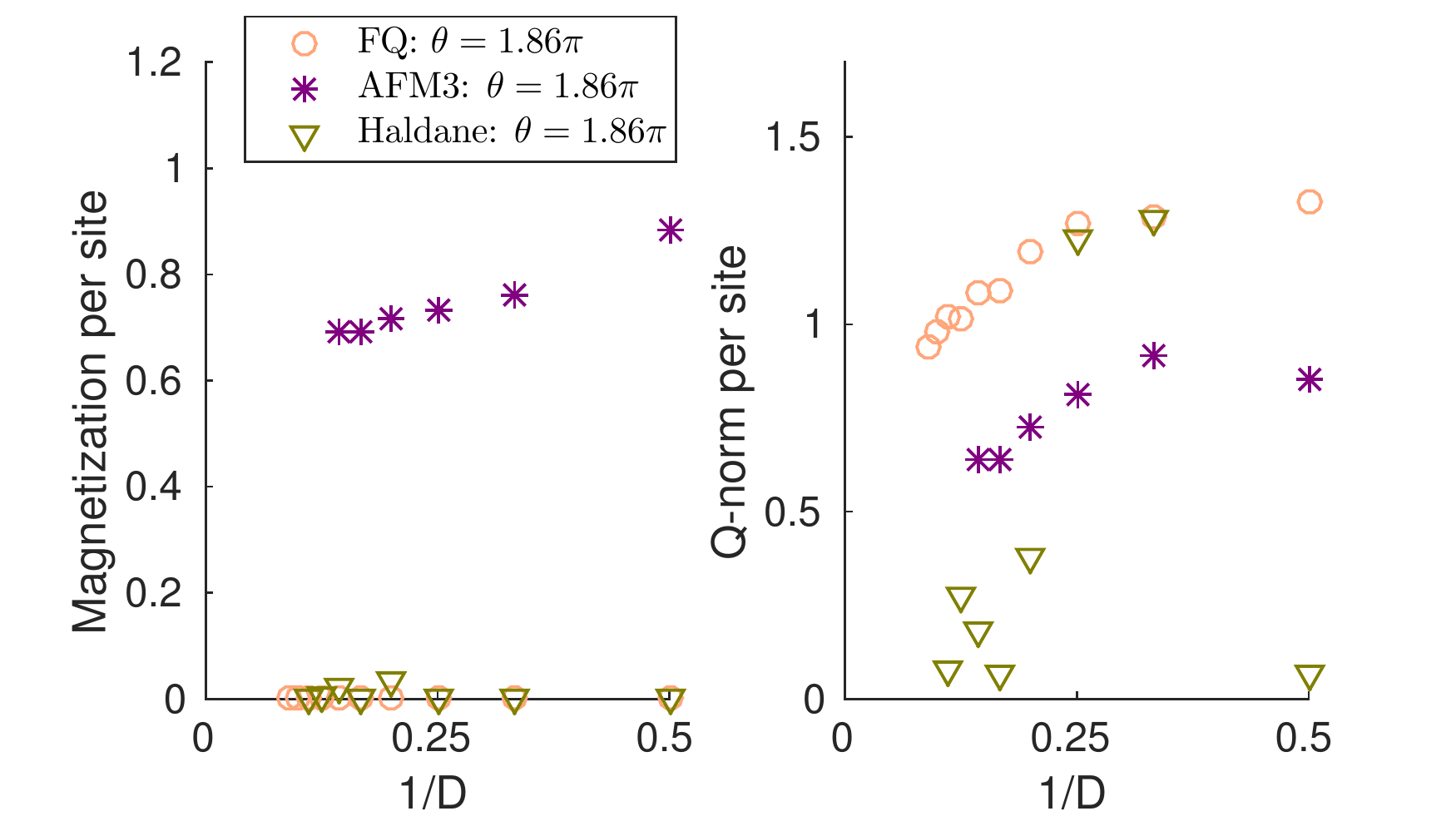}
	\caption{Energy (\emph{top}) and magnetization and Q-norm (\emph{bottom}) per site (full update) comparing the FQ and AFM3 simulations to simulations initialized in the Haldane phase at $\theta = 1.86\pi$. The fluctuations in both order parameters show that the Haldane-initialzed simulations do not stay in their original phase.}
	\label{fig-th186-hald}
	\end{figure}

	As a final check, we have initialized full update simulations from within the Haldane phase directly at the isotropic limit $J_{\text{anis}} = 1$ at two points of interest. Close to the FQ to AFM3 phase transition|at $\theta = 1.86\pi$|Fig.~\ref{fig-th186-hald} (top) shows that the simulations initialized in the Haldane phase are far from competitive, with the exception of the $D=3$ and $D=4$ simulations. However, Fig.~\ref{fig-th186-hald} (bottom) reveals that the aforementioned simulations are actually in the FQ phase, as their quadrupole moment shows. The other Haldane-initialized simulations also develop some quadrupolar order in the energy-minimization process, supporting the claim that the ground state is quadrupolar. Note that it therefore also does not make sense to do a $D\rightarrow \infty$ extrapolation on the Haldane-initialized simulations, as the simulations are not in a well-defined phase.

\begin{figure}[htb]
	\includegraphics[scale=0.5]{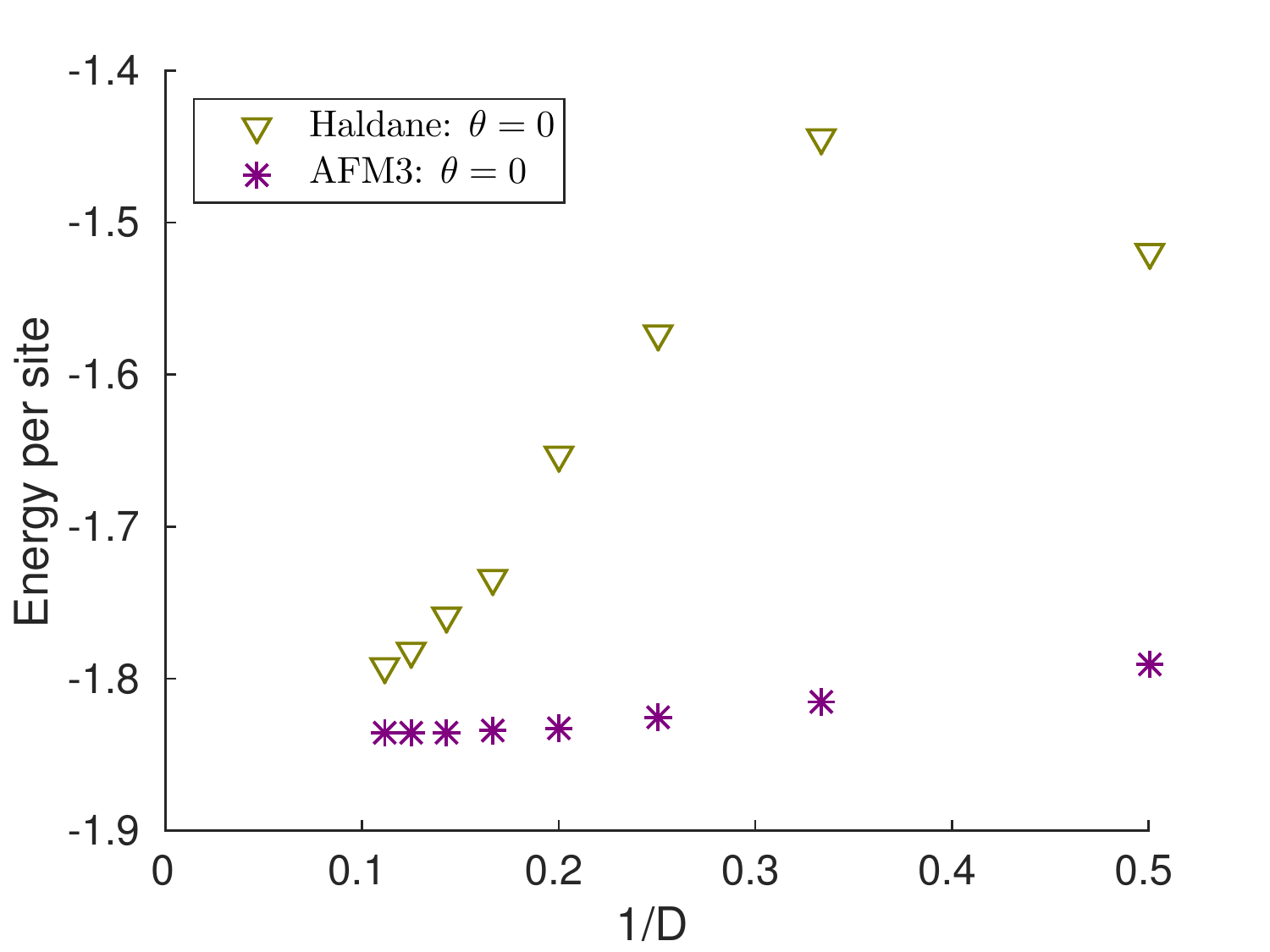}
	\includegraphics[scale=0.5]{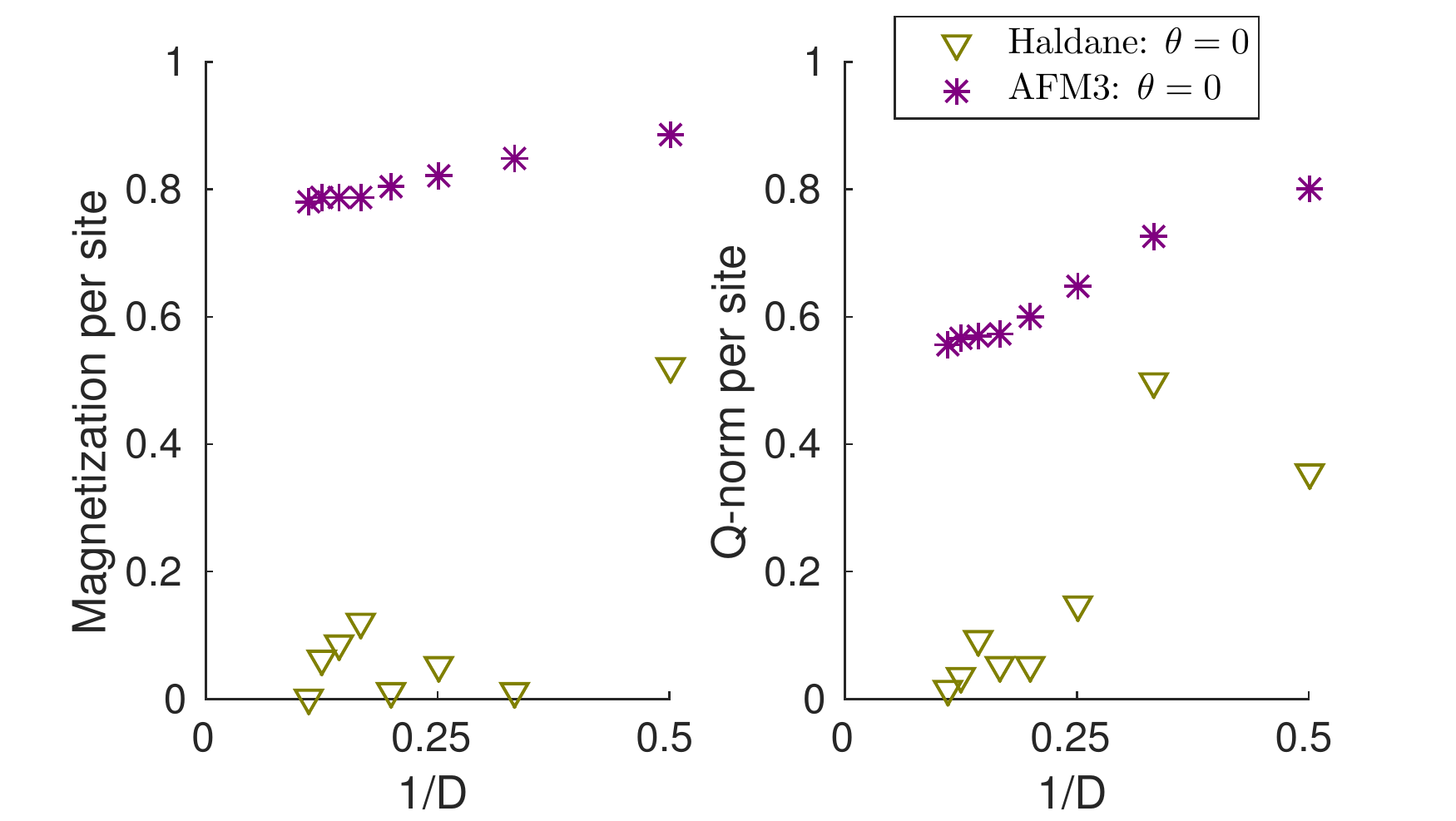}
	\caption{Energy (\emph{top}) and magnetization and Q-norm (\emph{bottom}) per site (full update) comparing the AFM3 simulations to simulations initialized in the Haldane phase at the Heisenberg point $\theta = 0$. Also here, the fluctuations in both order parameters show that the Haldane-initialized simulations do not stay in their original phase.}
	\label{fig-th0-hald}
	\end{figure}

	At the Heisenberg point $\theta = 0$, we encounter a similar situation. Also in this case, the energy of the Haldane-initialized simulations is far from competitive. Moreover, the Haldane-initialized simulations do not remain paramagnetic in the optimization process because they develop magnetic and quadrupolar order (Fig.~\ref{fig-th0-hald}). Thus, as before, it does not make sense to do a $D \rightarrow \infty$ extrapolation on the Haldane-initialized simulations.

	In conclusion: the ground state is ordered at both $\theta = 1.86\pi$ and $\theta = 0$, and the Haldane phase is absent in the isotropic limit.

\newpage

\bibliographystyle{apsrev4-1}
\bibliography{refs,bbs1t-notes}

\begin{thebibliography}{80}
\bibitem [{n-t()}]{n-threesubl}%
  \BibitemOpen
  \href@noop {} {}\bibinfo {note} {The ``3'' indicates three-sublattice order:
  it is used to differentiate 120$\degree$ magnetic order (AFM3) from ordinary
  antiferromagnetic order (AFM), the latter of which displays a two-sublattice
  pattern.}\BibitemShut {Stop}%
\bibitem [{n-i()}]{n-itegate}%
  \BibitemOpen
  \href@noop {} {}\bibinfo {note} {As mentioned in the previous paragraph, the
  imaginary time evolution gate $g$ used on the triangular lattice is actually
  a product of three single-bond gates (corresponding to a horizontal, vertical
  and diagonal bond that together form a closed triangle), rather than just a
  single-bond gate that is used on the square lattice.}\BibitemShut {Stop}%
\bibitem [{n-e()}]{n-errorbar}%
  \BibitemOpen
  \href@noop {} {}\bibinfo {note} {The extrapolations in this paragraph are all
  based on the (green) variational update energies. For the simple and full
  update points we can use the same procedure.}\BibitemShut {Stop}%
\bibitem [{n-d()}]{n-datapoints}%
  \BibitemOpen
  \href@noop {} {}\bibinfo {note} {The reason that the AFM3 data points for
  $D=8$ and $D=9$ at $\theta = 1.86\pi$ and $1.87\pi$ and $D=9$ at $\theta =
  1.88\pi$ are left out is because the corresponding simulations mixed with a
  ferroquadrupolar state (decreasing their magnetization significantly) during
  the optimization process, and are therefore not proper AFM3 simulations. Note
  that this does not happen the other way around because the imposed U(1)
  symmetry used for the FQ simulations is incompatible with the 120$\degree$
  magnetic order in the AFM3 phase. Regardless, the phase transition is clearly
  visible with the data points available.}\BibitemShut {Stop}%
\bibitem [{n-D()}]{n-D9bonds}%
  \BibitemOpen
  \href@noop {} {}\bibinfo {note} {There remains a difference in weights in $x$
  and $y$-bonds, but it shows mainly in the larger weights, not in the ones
  that are close to those being truncated.}\BibitemShut {Stop}%
\end{thebibliography}

\end{document}